\documentclass[12pt,preprint]{aastex}

\newcommand{\vect}[1]{\mbox{\boldmath$#1$}}
\newcommand{\dfrac}[2]{{\displaystyle \frac{#1}{#2}}  }
\newcommand{\alfven}{Alfv\'{e}n}
\newcommand{\eqref}[1]{(\ref{#1})}

\def\lesssim{\mathrel{\hbox{\rlap{\hbox{\lower4pt\hbox{$\sim$}}}\hbox{$<$}}}}
\def\gtrsim{\mathrel{\hbox{\rlap{\hbox{\lower4pt\hbox{$\sim$}}}\hbox{$>$}}}}

\slugcomment{ApJ Accepted}

\shorttitle{Viscous Disk-Planet Interaction}
\shortauthors{Muto and Inutsuka}

\begin{document}

\title{Local Linear Analysis of Interaction between a Planet and Viscous
Disk and an Implication on Type I Planetary Migration}

\author{Takayuki Muto\altaffilmark{1}}
\affil{Department of Physics, Kyoto University, \\
Kitashirakawa-oiwake-cho, Sakyo-ku, Kyoto, 606-8502, Japan}

\and

\author{Shu-ichiro Inutsuka}
\affil{Department of Physics, Nagoya University, \\
Furo-cho, Chikusa-ku, Nagoya, 464-8602, Japan}

\email{muto@tap.scphys.kyoto-u.ac.jp}
\altaffiltext{1}{JSPS Research Fellow}

\begin{abstract}
We investigate the effects of viscosity on disk-planet interaction
 and discuss how type I migration of planets is modified.
 We have performed a linear calculation using 
 shearing-sheet approximation and obtained the detailed, high resolution
 density structure around the planet embedded in a viscous disk with a
 wide range of viscous coefficients.
  We use a time-dependent formalism that is useful in investigating the
 effects of various physical processes on disk-planet interaction.  
  We find that the density structure
 in the vicinity of the planet is modified and the main contribution
 to the torque comes from this region, in contrast to inviscid case.
 Although it is not possible to derive total torque acting on the planet
 within the shearing-sheet approximation, the one-sided
 torque can be very different from the inviscid
 case, depending on the Reynolds number.  This effect has been neglected
 so far but our results indicate 
 that the interaction between a viscous disk and a planet can be 
 qualitatively different from an inviscid case and the details of the
 density structure in the vicinity of the planet is critically important.
\end{abstract}


\section{Introduction} \label{intro}

Disk-planet interaction and type I planetary migration is one of the
main issues in the theory of planet formation.  A low mass planet or the
core of the gas giant planet embedded in the protoplanetary disk excites
spiral density wave in the disk, which carries angular momentum flux
away from the planet.  The linear calculation of the spiral density wave
is first carried out by Goldreich and Tremaine (1979), and there has
been a number of extensive work assuming inviscid, isothermal disk such
as Ward (1986, 1989), Artymowicz (1993), and Tanaka et al. (2002). 
In the case of an inviscid, isothermal disk,
protoplanets migrate towards the central star as a result of disk-planet
interaction very rapidly, thereby imposing a serious problem on the
theory of planet formation.

The decay of the orbital semi-major axis of the protoplanet is a result
of the subtle difference between the inner disk (the part of the disk
inside 
the orbit of the protoplanet) and the outer disk (the part of the disk
outside the orbit of the protoplanet).  The outer disk exerts torque
on the protoplanet in such a way that the semi-major axis decreases, and
the inner 
disk exerts opposite torque.  In the lowest order of the local
approximation, or shearing-sheet approximation, these two effects cancel
each other, and the difference appears at higher order.  Therefore, type
I migration can be very sensitive to the physical state of the disk.
Indeed, when non-isothermal effect is taken into account, the
direction of migration can be reversed (Paardekooper and Mellema
2006, Baruteau and Masset 2008, Paardekooper and Papaloizou 2008).  It
is also discussed that the global magnetic field can slow down the rate
of type I migration or even reverse the
direction (Terquem 2003, Fromang et al. 2005,
Muto et al. 2008).

In order to construct a predictable, self-consistent model for planet
formation, it is important to make clear how different physical
processes of the 
disk affect disk-planet interaction and type I migration.
In this paper, we revisit the effects of viscosity on disk-planet
interaction.
There has been a number of work on this topic.  Meyer-Vernet and Sicardy
(1987) discussed that the 
viscosity does not affect the torque exerted on the planet as long as
its value is small, although the shape of the wake may be modified.
Papaloizou and Lin (1984, and a series of their paper) and Takeuchi et 
al. (1996) calculated that the
damping of the density wave and the transfer of the angular
momentum from the density wave to the background disk, resulting in the
gap formation around a massive planet.  Masset (2001) investigated
the effects of viscosity on the horseshoe regions and discussed how the
corotation torque may be altered by the effect of viscosity.
More recently, Paardekooper and Papaloizou (2009b) investigated the
dynamics of corotation region using both linear and non-linear
calculations.

Although there has been a number of investigations on the disk-planet
interaction in a viscous disk (for example, Masset 2001, 2002 or
Paardekooper and Papaloizou 2009a for corotation torque, D'Angelo et
al. 2002, 2003 for high resolution numerical study), 
there has not yet been a study of wide range in the viscous coefficient 
that requires an analysis on the detailed density
  structure in the vicinity of the planet.
  In this paper, as a first
step for the complete investigation, we show results of linear
calculation in a local, shearing-sheet analysis.
  We use time-dependent methods to calculate the disk response against
  the planet potential.  Although this method has not been widely used
  so far, this is useful in investigating the effects of
  various physical processes on disk-planet interaction.
  We calculate the
non-axisymmetric density structure around the planet and investigate how
the resulting torque is altered by the effect of shear viscosity.  We
have studied wide range of the parameters of viscous coefficient and
calculate the density structure with high resolution.  We find that the
density 
structure in the vicinity of the planet is altered in a viscous
disk, with viscous coefficient of $\sim 0.1$ in terms of $\alpha$ 
(standard $\alpha$ parameter, see
equation \eqref{alpha_def} for definition), which may
be realized as a result of the turbulence induced by magneto-rotational
instability (MRI, Balbus and Hawley 1991, Sano et al. 2004).  

Since it is not possible to calculate the total torque (differential
torque between the disk inside the orbit of the planet and outside) in
shearing-sheet approximation, we look at one-sided torque to investigate
how type I migration rate can be affected qualitatively by viscosity.
One-sided 
torque is the torque exerted by one side of the disk with respect to the
planet's orbital radius.  
We also note that shearing-sheet calculation 
is a useful first step before 
differential torque is calculated (Tanaka et al. 2002).
We find that as we increase the values of viscous coefficient $\alpha$, 
one-sided torque stays unchanged until $\alpha \lesssim 0.01$,
then increases until $\alpha \sim 1$, and finally decreases.  The torque
can be factor of two larger than inviscid case when $\alpha=0.1$, and
more importantly, the enhancement of the torque is a result of the
modified density structure in the vicinity of the planet, which has
not yet been investigated in detail.  Our results indicate that the
physical mechanisms of the disk-planet interaction in a viscous disk
may depend on the detailed density structure around the planet.

The plan of this paper is as follows.  In section \ref{basiceqn} we show
the basic equations and describe the time-dependent methods to obtain a
stationary, non-axisymmetric pattern of linear perturbation.  We also
discuss how two-dimensional (2D) and three-dimensional (3D) analyses are
related in this section.  In section \ref{results}, we show the results
of 2D and 3D calculations and show how the torque behaves as we vary the
amount of viscosity.  In section \ref{analyticdiscussion}, we explain
the qualitative behavior of the torque using a simple analytic model.  
Section \ref{discussionsummary} is for discussion and summary.

\section{Basic Equations and Numerical Methods}
\label{basiceqn}

\subsection{Linear Perturbation Analysis}
\label{linear_anal}

We consider isothermal Navier-Stokes equations with one
planet using shearing-sheet approximation.
\begin{eqnarray}
\label{EoC_full}
& \dfrac{\partial \rho}{\partial t} + \nabla\cdot(\rho \vect{v}) =
 0 \\ 
\label{EoM_full}
& \dfrac{\partial \vect{v}}{\partial t} + \vect{v}\cdot \nabla \vect{v}
 + 2 \Omega_{\rm p} \vect{e}_z \times \vect{v}
 = -\dfrac{c^2}{\rho} \nabla {\rho} + 3 \Omega^2_{\rm p} x
 \vect{e}_x + \nu \nabla^2 \vect{v} + \dfrac{1}{3}\nu
				      \nabla (\nabla \cdot
 \vect{v}) - \nabla \psi_{\rm p} 
\end{eqnarray}
where $\rho$ is density, $\vect{v}$ is velocity, 
$\Omega_{\rm p}$ is the Kepler angular velocity of the planet,
$\psi_{\rm p}$ is the gravitational potential of the planet, $\nu$ is
shear viscosity.  The third term of left hand side of equation
\eqref{EoM_full} is Coriolis force and the second term of right hand
side of equation \eqref{EoM_full} is tidal force.    
We assume the planet is located at the origin and
stationary with respect to this coordinate system, i.e.,
\begin{equation}
 \psi_{\rm p} = \psi_{\rm p} (x,y,z),
\end{equation}
where $\psi_{\rm p}$ does not depend on time.
In this paper, we neglect bulk viscosity for
simplicity.  We also neglect vertical stratification and assume that the
background density without a planet is homogeneous.  This gives an
uncertainty in the box size in the $z$-direction, which will be
discussed in Section \ref{Lzsize}.  We also neglect the effect of global
gas pressure gradient exerted on the background disk and assume that the
background gas is rotating at Kepler velocity.

The unperturbed state without a planet is given by
\begin{eqnarray}
& \rho = \rho_0 = \mathrm{const} \\
& \vect{v}_{0} = -\dfrac{3}{2} \Omega_{\rm p} x \vect{e}_y .
\end{eqnarray}
In the presence of viscosity, if we perform a global analysis, there is
a mass accretion onto the central 
star in general.  However, in the shearing-sheet approximation, where
linear background shear is assumed, this effect is not taken into
account.  We expect that the density structure only in the vicinity of
the planet can be well approximated even if we neglect the global mass
accretion.

We consider linear perturbation.  All the perturbation quantities are
denoted with $\delta$, e.g., $\delta \rho$ for density perturbation.  
Perturbation equations are given by 
\begin{equation}
 \left(\dfrac{\partial}{\partial t} - \dfrac{3}{2} \Omega_{\rm p} x
  \dfrac{\partial}{\partial y}\right) \dfrac{\delta
 \rho}{\rho_0} + \nabla\cdot\delta\vect{v} = 0
 \label{EoC_pert}
\end{equation}
\begin{eqnarray}
& \left(\dfrac{\partial}{\partial t} - \dfrac{3}{2} \Omega_{\rm p} x
  \dfrac{\partial}{\partial y}\right) \delta \vect{v} 
 - 2\Omega_{\rm p} \delta v_y \vect{e}_x
 + \dfrac{1}{2} \Omega_{\rm p} \delta v_x \vect{e}_y  \nonumber \\
& = - c^2 \nabla
 \dfrac{\delta \rho}{\rho_0} 
 + \nu \nabla^2 \delta \vect{v}
 + \dfrac{1}{3} \nu \nabla \left( \nabla \cdot \delta \vect{v} \right) 
 - \nabla \psi_{\rm p} 
 \label{EoM_pert}
\end{eqnarray}
We solve equations \eqref{EoC_pert} and \eqref{EoM_pert} to obtain a
steady state solution and calculate torque exerted on one side (either
$x>0$ or $x<0$) of the disk by the planet.  The torque exerted on the
planet is obtained as a backreaction of this torque.  
Since we use shearing-sheet approximation, torque exerted from each side
of the planet is the same in magnitude and opposite in sign.  Although
we do not obtain a net torque, it is still possible to investigate how
the disk structure is affected by the viscosity and qualitatively
predict how the disk and planet interact.

In order to obtain a steady state solution, 
we solve linear perturbation equations \eqref{EoC_pert} and
\eqref{EoM_pert} using Fourier transform methods given by Goodman and
Rafikov (2001).
  We transform the equations into the shearing coordinate
  $(t^{\prime}, x^{\prime}, y^{\prime}, z^{\prime})$ defined by
\begin{eqnarray}
&& t^{\prime} = t \\
&&  x^{\prime} = x \\
&& y^{\prime} = y + \dfrac{3}{2} \Omega_{\rm p} x t \\
&& z^{\prime} = z.
\end{eqnarray}
In this coordinate system, temporal and spatial derivatives are given by 
\begin{equation}
 \dfrac{\partial}{\partial t} =
  \dfrac{\partial}{\partial t^{\prime}} + \dfrac{3}{2} \Omega_{\rm p}
  x^{\prime} \dfrac{\partial}{\partial y^{\prime}},
  \label{dt_shear}
\end{equation}
\begin{equation}
 \dfrac{\partial}{\partial x} = 
  \dfrac{\partial}{\partial x^{\prime}} + \dfrac{3}{2} \Omega_{\rm p}
  t^{\prime} \dfrac{\partial}{\partial y^{\prime}},
  \label{dx_shear}
\end{equation}
\begin{equation}
 \dfrac{\partial}{\partial y} = \dfrac{\partial}{\partial y^{\prime}}, 
 \ 
 \dfrac{\partial}{\partial z} = \dfrac{\partial}{\partial z^{\prime}} .
 \label{dydz_shear}
\end{equation}
Then, perturbation equations become
\begin{equation}
 \dfrac{\partial}{\partial t^{\prime}} \dfrac{\delta
 \rho}{\rho_0} + \nabla \cdot \delta \vect{v} = 0
 \label{EoC_pert_shear}
\end{equation}
\begin{eqnarray}
& \dfrac{\partial}{\partial t^{\prime}} \delta \vect{v} 
 - 2\Omega_{\rm p} \delta v_y \vect{e}_x
 + \dfrac{1}{2} \Omega_{\rm p} \delta v_x \vect{e}_y  \nonumber \\
& = - c^2 \nabla
 \dfrac{\delta \rho}{\rho_0} 
 + \nu \nabla^2 \delta \vect{v}
 + \dfrac{1}{3} \nu \nabla \left( \nabla \cdot \delta \vect{v} \right) 
 - \nabla \psi_{\rm p} ,
 \label{EoM_pert_shear}
\end{eqnarray}
where the spatial derivatives in $\nabla$ are given by equations
\eqref{dx_shear} and \eqref{dydz_shear}.

  The coefficients of equation \eqref{EoC_pert_shear} and
  \eqref{EoM_pert_shear} are now independent of
  $(x^{\prime},y^{\prime},z^{\prime})$.  Therefore, if we Fourier
  transform in spatial directions, we obtain a set of
  ordinary differential equations decoupled for each
  $(k_x^{\prime},k_y^{\prime},k_z^{\prime})$ mode (Goldreich and
  Lynden-Bell 1965),
\begin{equation}
 \delta f(t^{\prime},x^{\prime},y^{\prime},z^{\prime}) 
  = \sum \delta f (t^{\prime}, k_x^{\prime}, k_y^{\prime}, k_z^{\prime}) 
  \exp \left[ i(k_x^{\prime} x^{\prime} + k_y^{\prime} y^{\prime} +
	k_z^{\prime} z^{\prime}) \right].
  \label{decomposition_FT}
\end{equation}
The relationship of wavenumber in $(x,y,z)$-coordinate and
$(x^{\prime},y^{\prime},z^{\prime})$-coordinate is given by
\begin{equation}
 k_x(t) = k_{x}^{\prime} + \dfrac{3}{2} \Omega_{\rm p} k_y t.
  \label{kx_evol}
\end{equation} 
\begin{equation}
 k_y = k_y^{\prime},\  k_z = k_z^{\prime}
\end{equation} 
Equation \eqref{kx_evol} indicates that the value of radial wavenumber
in $(x,y,z)$-coordinate evolves with time owing to the background
shear.  The value of radial wavenumber in shearing coordinate
$k_x^{\prime}$ gives the initial value of radial wavenumber in
$(x,y,z)$-plane.    
Equations of continuity \eqref{EoC_pert} and motion \eqref{EoM_pert} are
now 
\begin{equation}
 \dfrac{d}{dt} \dfrac{\delta \rho}{\rho_0} + ik_x(t) \delta v_x + ik_y
  \delta v_y + ik_z \delta v_z = 0,
  \label{EoC_pert_four}
\end{equation}
\begin{eqnarray}
&& \dfrac{d}{dt} \delta v_x - 2\Omega_{\rm p} \delta v_y =
  -c^2 ik_x(t)
  \dfrac{\delta \rho}{\rho_0} - \nu (k_x(t)^2+k_y^2+k_z^2) 
  \delta v_x \nonumber \\
&& - \dfrac{1}{3} \nu k_x(t) (k_x(t) \delta v_x + k_y \delta v_y + k_z
  \delta v_z) - ik_x(t) \psi_{\rm p},
  \label{EoMx_pert_four}
\end{eqnarray}
\begin{eqnarray}
&& \dfrac{d}{dt} \delta v_y + \dfrac{1}{2}\Omega_{\rm p} \delta v_x 
  = -c^2 ik_y \dfrac{\delta \rho}{\rho_0} 
  - \nu (k_x(t)^2+k_y^2+k_z^2) \delta v_y  \nonumber \\
&&  - \dfrac{1}{3} \nu k_y (k_x(t) \delta v_x + k_y \delta v_y + k_z
  \delta v_z) - ik_y \psi_{\rm p},
  \label{EoMy_pert_four}
\end{eqnarray}
and
\begin{eqnarray}
&& \dfrac{d}{dt} \delta v_z  
  = -c^2 ik_z \dfrac{\delta \rho}{\rho_0} 
  - \nu (k_x(t)^2+k_y^2+k_z^2) \delta v_y \nonumber \\
&&  - \dfrac{1}{3} \nu k_y (k_x(t) \delta v_x + k_y \delta v_y + k_z
  \delta v_z) - ik_z \psi_{\rm p} .
  \label{EoMz_pert_four}
\end{eqnarray}
In equations \eqref{EoC_pert_four}-\eqref{EoMz_pert_four}, we write all
the terms in $(t,x,y,z)$-coordinate.  These equations describe the
excitation and evolution of density wave by the source 
$\psi_{\rm p}$.  The source of the density wave becomes zero when
wavenumber $(k_x^{\prime},k_y^{\prime},k_z^{\prime})$ is large.

  As an initial condition, we assume that there is no
perturbation: $\delta \rho(t=0) = \delta \vect{v}(t=0) = 0$,
 and $k_{x}^{\prime}$ is taken to be sufficiently large in absolute
 magnitude and negative (positive) in sign for positive
 (negative) $k_y$.  If we use spatial resolution of $x$-direction
 $\Delta x$, we should take $k_{x}^{\prime}= \mp 2\pi/\Delta x$, where
 upper 
 (lower) sign is used for positive (negative) $k_y$.
As a result of time evolution, $k_x(t)$ increases (decreases) for each
positive (negative) $k_y$ mode.  When $k_x(t)$ reaches 
$\pm 2\pi / \Delta x$, where upper (lower) sign is for positive
(negative) $k_y$ modes, 
 we obtain a steady state profile of perturbation
quantities in Fourier space for non-axisymmetric ($k_y \neq 0$) modes.
\footnote{
The reason why we obtain a steady state as a result of time evolution is
as follows.  Let us assume that if $|k_x|>K_c$, the source $\psi_{\rm p}$
is so small that we can approximate it to be zero.  
The Fourier amplitude of the specific
$k_{x0}$ mode in $(t,x,y,z)$-coordinate at time $t=t_0$ is given by the
solution at $t=t_0$ with 
$k_x^{\prime}=k_{x0}-(3/2)\Omega_{\rm p} k_y t_0$ mode.  Therefore, the
absolute magnitude of 
$k_x^{\prime}$ that contributes to the Fourier amplitude of fixed
$k_{x0}$ becomes large as time $t_0$ increases.  If the absolute
magnitude of 
$k_x^{\prime}$ is larger than $K_c$, there is no wave excitation for this
mode until $|k_x(t)|$ becomes smaller than $K_c$.  Therefore, when
$|k_x^{\prime}|>K_c$, the contribution to the $k_{x0}$ mode is always
the same regardless of $t_0$.
}
  The Fourier amplitude of each $k_x$ mode
in steady state is given by the time evolution data through equation
\eqref{kx_evol}.  
The profile of physical quantities in Fourier space is then inverse
Fourier transformed to obtain values in real space.
Since we are interested in torque, we do not calculate axisymmetric
modes ($k_y=0$).

The standard procedure to obtain steady state solution is as follows.
 Firstly, stationary solution in the frame corotating with the planet is
 assumed: $\partial / \partial t = 0$.  Then, Fourier transformation in
 the $y$- and $z$-directions is performed to obtain the ordinary
 differential equations in the $x$-direction.  Finally, these ordinary
 differential equations are solved by imposing outgoing boundary
 condition, or equivalently the 
boundary condition that admits only  trailing wave, see e.g., 
Goldreich and Tremaine (1979), Korycansky and Pollack (1993),
 or Tanaka et al. (2002).  

In the presence of viscosity, this method
 introduces higher order derivative with respect to $x$, since viscous
 terms include the second order derivative in the radial direction.  The 
 resulting ordinary differential equations are higher order in the
 $x$-derivative than in inviscid cases. 
 The terms with the highest order derivative comes from viscous terms
 and their coefficients are viscous coefficient $\nu$.  
  In this case, it is difficult to take natural
 $\nu\to 0$ limit, since equations become singular in this limit.

  Time-dependent approach overcomes this difficulty
 since the order of time derivative is not affected by viscous terms
 since they do not include any time derivative.
 It is also easy to take $\nu\to 0$ limit since viscous
 terms simply drop from equations  
 \eqref{EoMx_pert_four}-\eqref{EoMz_pert_four} in this limit 
 and they do not introduce any singularity.

 In our formulation using the Fourier transform in sheared coordinate,
 the outgoing boundary condition in the radial direction 
 that is assumed in the standard stationary 
 formulation is not exactly satisfied, since Fourier 
 transform introduces the periodic boundary condition in the sheared 
 coordinate.  However, we find that outgoing boundary condition is 
 effectively satisfied if we vary the box size in the $x$-direction
depending on the modes 
 specified by $k_y$ and $k_z$.  Details of our methods are given below.  

 We also note that if we assume an initial condition with non-zero
 perturbation, additional homogeneous wave is introduced.  
 This wave has both leading
 ($k_x k_y<0$)  and trailing ($k_x k_y>0$) components and
 the resulting solution
 is simply the superposition of the specific solution of Equations
 \eqref{EoC_pert_four}-\eqref{EoMz_pert_four} assuming zero initial
 condition and homogeneous ($\psi_{\rm p}=0$) solution assuming the
 specified initial condition. 
The solution with non-zero initial condition is
 irrelevant in the present problem because 
 we consider the perturbation that is induced by the gravitational
 perturbation by the planet.

We write the box size in $(x,y,z)$-directions by $(L_x,L_y,L_z)$ and the
coordinate system extends $-L_x/2<x<L_x/2$ and so forth for other
directions.  
Practically, our procedure to obtain the stationary, non-axisymmetric
structure of the disk is summarized as follows.
\begin{enumerate}
 \item We take $k_x^{\prime}=\mp2\pi/\Delta x$ for positive (negative)
       $k_y$ modes.
 \item Equations \eqref{EoC_pert_four}-\eqref{EoMz_pert_four} are solved
       with initial condition 
       $\delta \rho (t=0)=\delta \vect{v}(t=0)=0$.
 \item Resulting solutions in $(k_x,k_y,k_z)$ space are inverse Fourier
       transformed to real space.
\end{enumerate}
In steps 2 and 3, there are some points we need to take care.  
In calculating the Fourier
modes, we have varied the mesh number in $k_x$ directions, which
corresponds to the step size of time, in such a way that all the
oscillations are well resolved for each $(k_y,k_z)$ mode.  The mesh
number in $k_x$ direction becomes as much as $10^6$ for small $k_y$.  
 We perform inverse Fourier transform in $k_x$ direction using the
 increased number of mesh in $k_x$ direction.
This procedure effectively makes the box size of $x$-direction longer
for a specific $k_y$ mode, and the data corresponding to necessary $x$, 
which is $-L_x/2<x<L_x/2$, is then interpolated from the resulting
profile in $(x,k_y,k_z)$ space.  This data is then used to perform
inverse Fourier transform in $(k_y,k_z)$ directions.  In this way, it is
possible to avoid aliasing effect in $x$-direction
 and outgoing boundary
condition in the steady state solution is effectively satisfied.
We do not use any window function, which has been incorporated by
Goodman and Rafikov (2001), in Fourier transform.  We have checked
that the resulting torque is not affected by the window function in the
absence of viscosity.  Also, since
window function introduces additional artificial dissipation, this
affects our results with viscosity.  
  
Once we have obtained the profiles of perturbation quantities in real
space, we can calculate the one-sided torque exerted on $x>0$ part of
the disk by the planet by
\begin{equation}
 T = - r_{\rm p} \int_{0}^{L_x}\int_{-L_y/2}^{L_y/2}
  \int_{-L_z/2}^{L_z/2} dxdydz  
  \delta \rho (x,y,z) \dfrac{\partial \psi_{\rm p}}{\partial y},
  \label{def_totaltorque}
\end{equation}
where $r_p$ is the semi-major axis of the planet, 
and the torque exerted on the planet is obtained as backreaction. 
For later convenience, we define ``torque distribution'' $T(x,y,z)$
that is simply the torque exerted on the fluid element located at
$(x,y,z)$,
\begin{equation}
 T(x,y,z) = -r_{\rm p} \delta \rho (x,y,z) 
 \dfrac{\partial \psi_{\rm p}}{\partial y}
 \label{def_tqdist}
\end{equation}
and ``torque density'' that is the torque exerted on an annulus of the
disk at $x$,
\begin{equation}
 T(x) = -r_{\rm p} \int_{-L_y/2}^{L_y/2} \int_{-L_z/2}^{L_z/2} dzdy
  \delta \rho (x,y,z) \dfrac{\partial \psi_{\rm p}}{\partial y}
  \label{def_tqdens}
\end{equation}

We normalize equations using $c$,
$\Omega_{\rm p}$, and 
$\rho_0$ so that the homogeneous equations ($\psi_{\rm p}$=0) contain
only one dimensionless variable $\alpha$ that is defined by
\begin{equation}
 \alpha = \dfrac{\nu}{cH},
 \label{alpha_def}
\end{equation}
where $H=c/\Omega_{\rm p}$ is the scale height of the disk.
Since we consider the linear perturbation analysis, the amplitude of the
perturbation is proportional to the normalized planet mass:
\begin{equation}
 \mu = GM_{\rm p}/Hc^2.
\end{equation}
We show the results with $\mu=1$ in subsequent sections.

We use the fifth order Runge-Kutta method to solve time evolution and
FFT routine given by Press et al. (1992) to perform Fourier transform.  
We use box size of $L_x=10H$ and $L_y=40H$, and the grid number in $x$- 
and $y$-directions $(N_x,N_y)=(512,512)$.
The three-dimensional results depend on the box size in $z$-direction,
$L_z$, which is discussed in the next subsection.

\subsection{Vertical Box Size and the Difference between 2D and 3D
  Calculation}
\label{Lzsize}

In this subsection, we discuss how 2D and 3D calculations differ from
each other and determine the relevant box size in the $z$-direction with 
which the effect of vertical stratification is effectively taken into
account. 

\begin{figure}
 \plotone{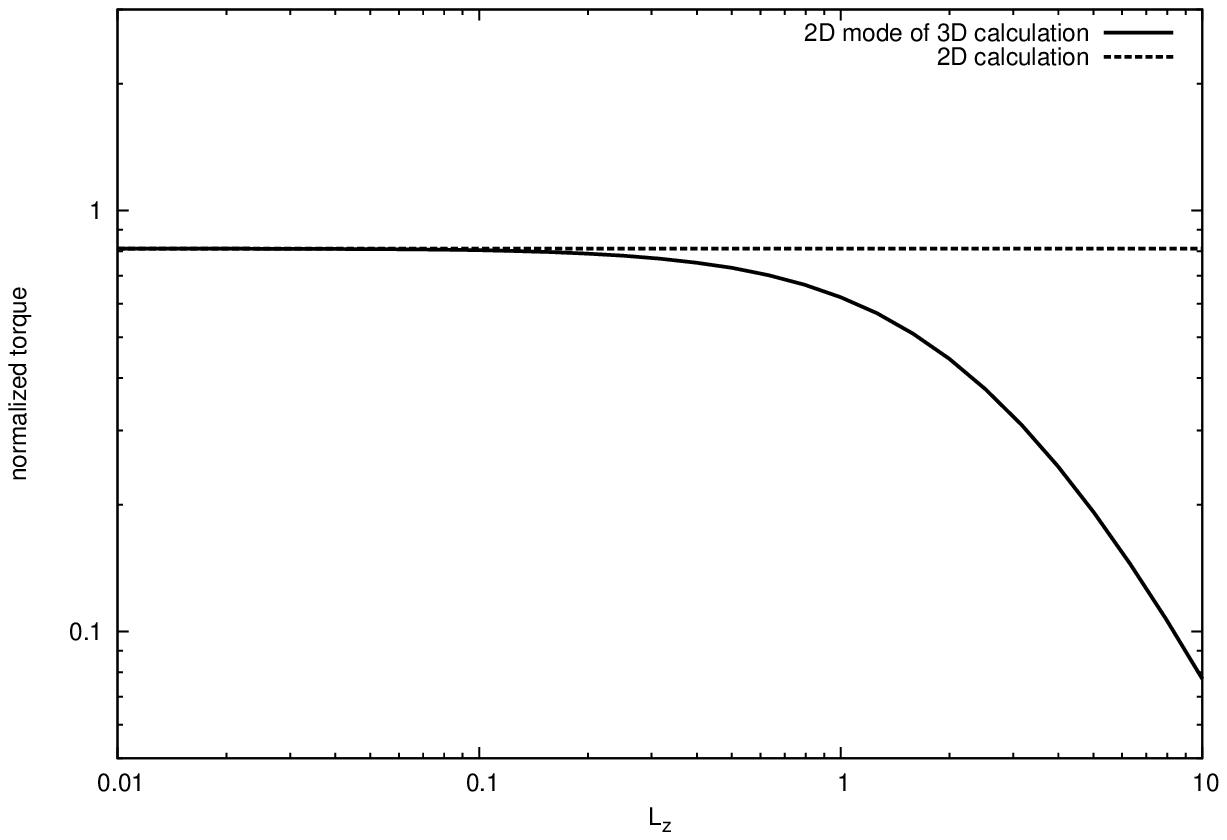}
 \caption{Comparison of the torque obtained in 2D calculation and the 2D
 mode of 3D calculation for various $L_z$.  In 2D calculations (dashed
 line), we use the gravitational potential given by equation
 \eqref{psi_pure2D}, while in 2D mode of 3D calculation (solid line),
 we use the gravitational potential given by equation
 \eqref{psi_unstrat}.  The value of the normalized torque,
 $T/\Gamma_{\rm 2D}$ is plotted, where $\Gamma_{\rm 2D}$ is defined by
 equation \eqref{torque_norm_2D}.}
 \label{fig:varheight}
\end{figure}

We compare 2D calculation and 2D mode of 3D calculation.  By ``2D
calculation'', we mean the gravitational potential of the planet is
given by 
\begin{equation}
 \psi_{\rm p,2D} (x,y)
  = - \dfrac{GM_{\rm p}}{\sqrt{x^2 + y^2 + \epsilon_{\rm 2D}^2}},
  \label{psi_pure2D}
\end{equation}
where $\epsilon_{\rm 2D}$ denotes the softening length of 2D
calculation, which is usually incorporated in most of 2D work.  
By ``2D mode of 3D calculation'', we mean that the analysis is
restricted to $k_z=0$ mode of 3D calculation.
  Three-dimensional potential is given by
\begin{equation}
 \psi_{\rm p,3D} (x,y,z) 
  = -\dfrac{GM_{\rm p}}{\sqrt{x^2 + y^2 + z^2 + \epsilon_{\rm 3D}^2}}. 
  \label{psi_full3D}
\end{equation}
  Therefore, gravitational potential for ``2D mode of 3D calculation''
  is given by  
\begin{eqnarray}
& \psi_{\rm p,3D} (x,y)
 &= - \dfrac{1}{L_z} \int_{-L_z/2}^{L_z/2} dz
  \dfrac{G M_{\rm p}}{\sqrt{x^2 + y^2 + z^2 + \epsilon_{\rm 3D}^2}}
  \nonumber \\ 
 && = -\dfrac{GM_{\rm p}}{L_z} \log
  \left| \dfrac{x^2 + y^2 + \epsilon_{\rm 3D}^2 + L_z^2/2 + L_z
   \sqrt{L_z^2/4 + x^2 + y^2 + \epsilon_{\rm 3D}^2}}{x^2 + y^2 +
   \epsilon_{\rm 3D}^2} \right|,
  \label{psi_unstrat}
\end{eqnarray}
where we denote softening length in 3D calculation by 
$\epsilon_{\rm 3D}$.  We note that $\psi_{\rm p,2D}$ and
 $\psi_{\rm p,3D}$ coincides if $x^2+y^2+\epsilon^2 \gg L_z$, but they
 differ if $x^2+y^2+\epsilon^2 \ll L_z$, since in this case, 3D
 potential behaves as $\psi_{\rm p,3D} \sim \log r$
 while 2D potential behaves as $\psi_{\rm p,2D} \sim 1/r$,
 where $r^2=x^2+y^2+\epsilon^2$.
  Vertical averaging makes the potential weaker in the vicinity of the
  planet, thereby producing small  
perturbation in 2D mode of 3D calculation.
  Figure \ref{fig:varheight} shows the torque obtained by
2D inviscid ($\nu=0$) calculation using $\psi_{\rm p,2D}$ and
$\psi_{\rm p,3D}$ with various $L_z$.  We note that in producing Figure 
\ref{fig:varheight}, one-sided torque is calculated according to
equation \eqref{def_totaltorque} and the value is normalized by
\begin{equation}
 \Gamma_{\rm 2D} = \mu^2 \rho_0 L_z r_{\rm p} H c^2 .
  \label{torque_norm_2D}
\end{equation}
It is clear that 2D and 3D results coincide if $L_z$ is small, but the
3D result becomes small when $L_z$ is large.

\begin{figure}
 \plotone{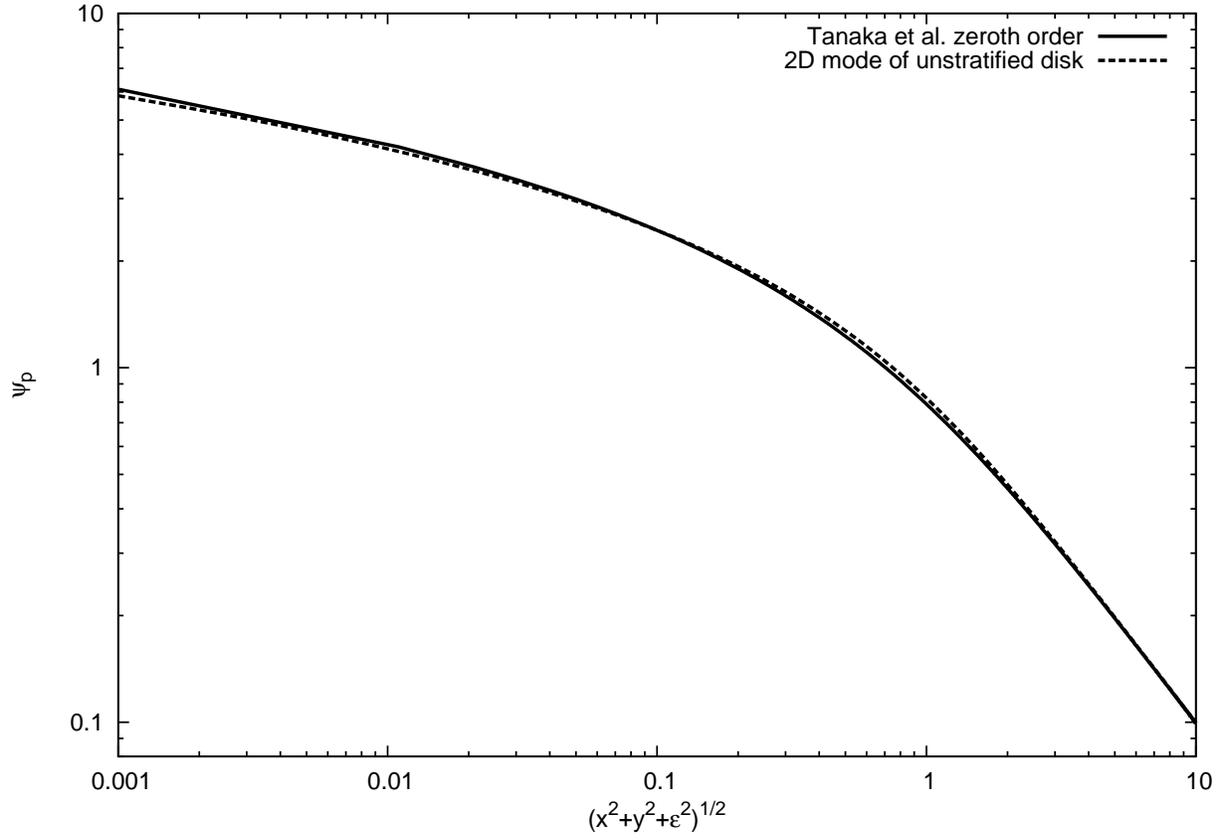}
 \caption{Comparison of the 2D mode of the potential used by Tanaka et
 al. (2002) given by equation \eqref{psi_strat} (solid line) and the
 $k_z=0$ mode of the potential given by \eqref{psi_unstrat} 
with $L_z=2.7H$ (dashed line).} 
 \label{fig:potfit}
\end{figure}

The reasonable value of $L_z$ may be obtained by comparing the results
of 2D mode of 3D calculation with calculation in which vertical
stratification is 
taken into account.  Tanaka et al. (2002) performed such
calculations for an inviscid disk.  Their method is to decompose
vertical modes in terms 
of Hermite polynomials, and they have found that the vertically averaged 
2D mode gives most of the contribution to the resulting torque.  Their
2D mode of the potential, $\psi_{\rm p,TTW}$, is given by
\begin{eqnarray}
& \psi_{\rm p,TTW} &= - \dfrac{1}{\sqrt{2\pi}H} \int_{-\infty}^{\infty} dz
  \dfrac{GM_{\rm p}}{\sqrt{x^2+y^2+z^2+\epsilon_{\rm 3D}^2}} 
  e^{-z^2/2H^2} \nonumber \\
 &&= - \dfrac{GM_{\rm p}}{\sqrt{2\pi}H}
 \exp \left[ (x^2+y^2+\epsilon_{\rm 3D}^2)/4 \right] 
  K_0 \left[ \left( x^2+y^2+\epsilon_{\rm 3D}^2 \right)/4 \right],
  \label{psi_strat}
\end{eqnarray}
where $K_0(x)$ is the modified Bessel function of the zeroth order.
Fitting  
equation \eqref{psi_unstrat} with equation \eqref{psi_strat}, we find
$L_z=2.7H$ gives a reasonable agreement, see Figure \ref{fig:potfit}.
Zeroth order of modified local
approximation incorporated by Tanaka et al. (2002) coincides with the
shearing-sheet approximation, and we have exactly the same homogeneous
equations with them for 2D ($k_z=0$) mode
 if we assume $\partial/\partial t=0$.  As is clear from Figure
\ref{fig:varheight}, 2D mode in 3D calculation gives smaller amount of
torque than the calculation using 2D potential given by equation
\eqref{psi_pure2D}.  This explains why Tanaka et al. (2002) has obtained 
slightly smaller amount of torque compared to 2D results. 
We have also checked that setting $L_z=2.7H$, our results of one-sided
torque agree with those determined by using Tanaka et al. (2002) methods
within an error of $3 \%$.
\footnote{
There is another complication regarding the normalization of the
torque when comparing the value of the torque obtained by our
calculation and that given by Tanaka et al. (2002).  They
use surface density to normalize the torque they have
obtained.   We use normalization given by equation
\eqref{torque_norm_2D}.  Correspondence between the two results is
given by reading our $\rho_0 L_z$ to $\sigma_{\rm p}$ of Tanaka et
al. (2002), and this is how we have obtained the agreement with their
calculation. 
}

We use $L_z=2.5H$ and the mesh number
in $z$-direction is taken to be $128$.  In total, we have performed
calculation with $(L_x, L_y, L_z)=(10H,40H,2.5H)$ 
with mesh number $(512,512,128)$, and the resulting mesh size is 
$(\Delta x, \Delta y, \Delta z)=(0.02H,0.08H,0.02H)$. 
However, the radial box size is variably
extended according to modes in calculation, as discussed in Section
\ref{linear_anal}.  Effective box size in 
$x$-direction is as much as $10^4 H$.
We use $\epsilon_{\rm 3D}=10^{-3}H$ for the softening parameter.
Our results vary upto $30 \%$ for large values of viscosity when
smoothing length is varied upto twice 
the grid resolution.  Therefore, our results give at least a qualitative
view of how disk-planet interaction is altered by the effects of
viscosity.  The variation of one-sided torque as a function of smoothing
parameter is further discussed in Section \ref{results:3D}.

\section{Results}
\label{results}

In this section, we show our results of density structure and one-sided
torque obtained by linear analysis.  In this section, the normalization
of the torque is given by
\begin{equation}
 \Gamma_{\rm 3D} = \mu^2 \rho_0 r_{\rm p} H^2 c^2.
\end{equation}
Note the difference in the normalization from that used in Section
\ref{Lzsize}.  The results of 2D mode torque given in Section
\ref{restrict_2D} are 
different by a factor of $L_z/H=2.5$ from those given in Figure
\ref{fig:varheight}.  The normalization of torque distribution defined
by equation \eqref{def_tqdist} is taken to be
\begin{equation}
 \dfrac{T(x,y,z)}{\mu^2 \rho_0 r_{\rm p} c^2 H^{-1}},
  \label{norm_tqdist}
\end{equation}
and torque density $T(x)$ defined by equation \eqref{def_tqdens} is
given by
\begin{equation}
 \dfrac{T(x)}{\mu^2 \rho_0 r_{\rm p} c^2}.
  \label{norm_tqdens}
\end{equation}

\subsection{Calculations Restricted to 2D Mode}
\label{restrict_2D}

\begin{figure}
 \plotone{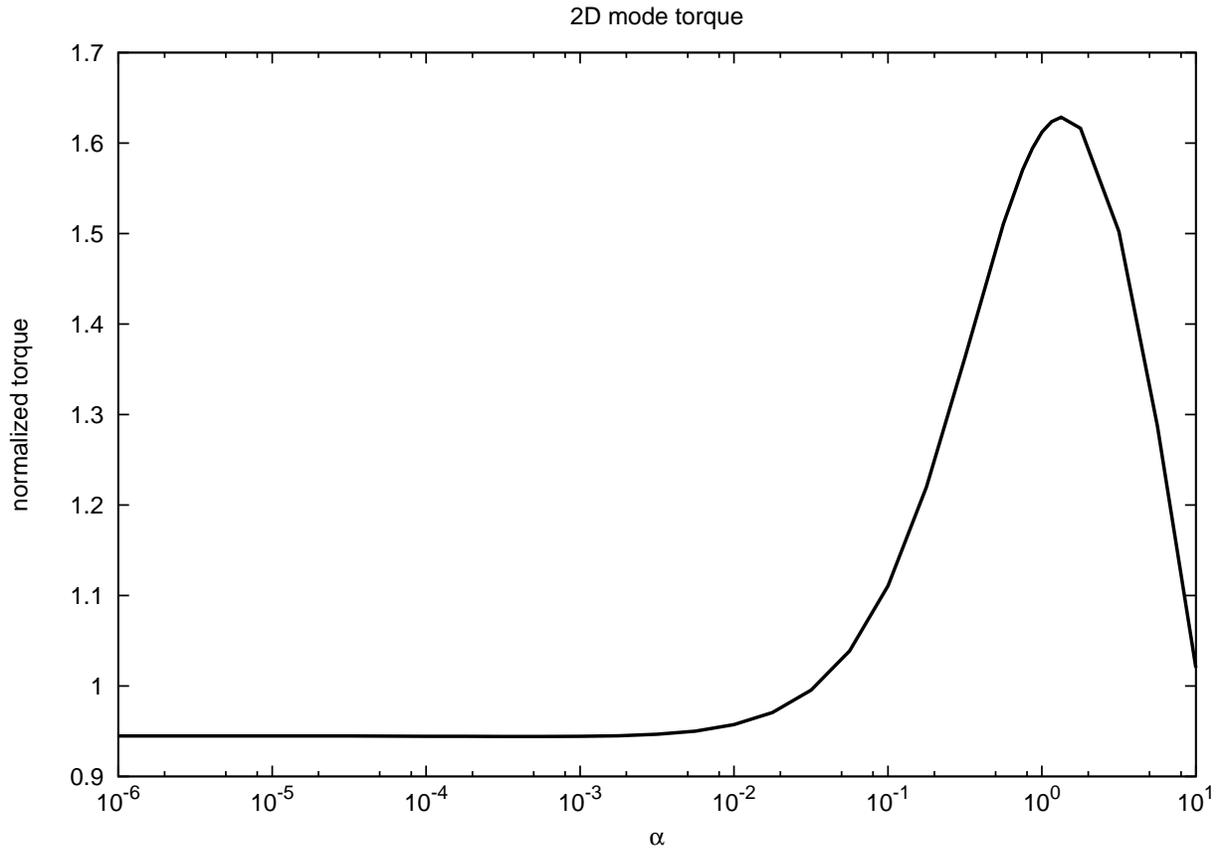}
 \caption{Variation of torque as a function of viscous coefficient
 $\alpha$ as a result of 3D calculation restricted to 2D modes.} 
 \label{fig:2D_torque}
\end{figure}

We first show the results of calculations restricted to 2D modes
($k_z=0$). 
Although this is only an approximation, physics involved is made clear.
Figure \ref{fig:2D_torque} shows the torque obtained for various
viscosity parameter $\alpha$.  For small viscous coefficients 
($\alpha \lesssim 10^{-2}$), torque is not affected by the viscosity, as
discussed by Meyer-Vernet and Sicardy (1987).  However, when viscous
coefficient is large, it is shown that one-sided torque increases and
peaks around $\alpha \sim 1$.

\begin{figure}
 \plotone{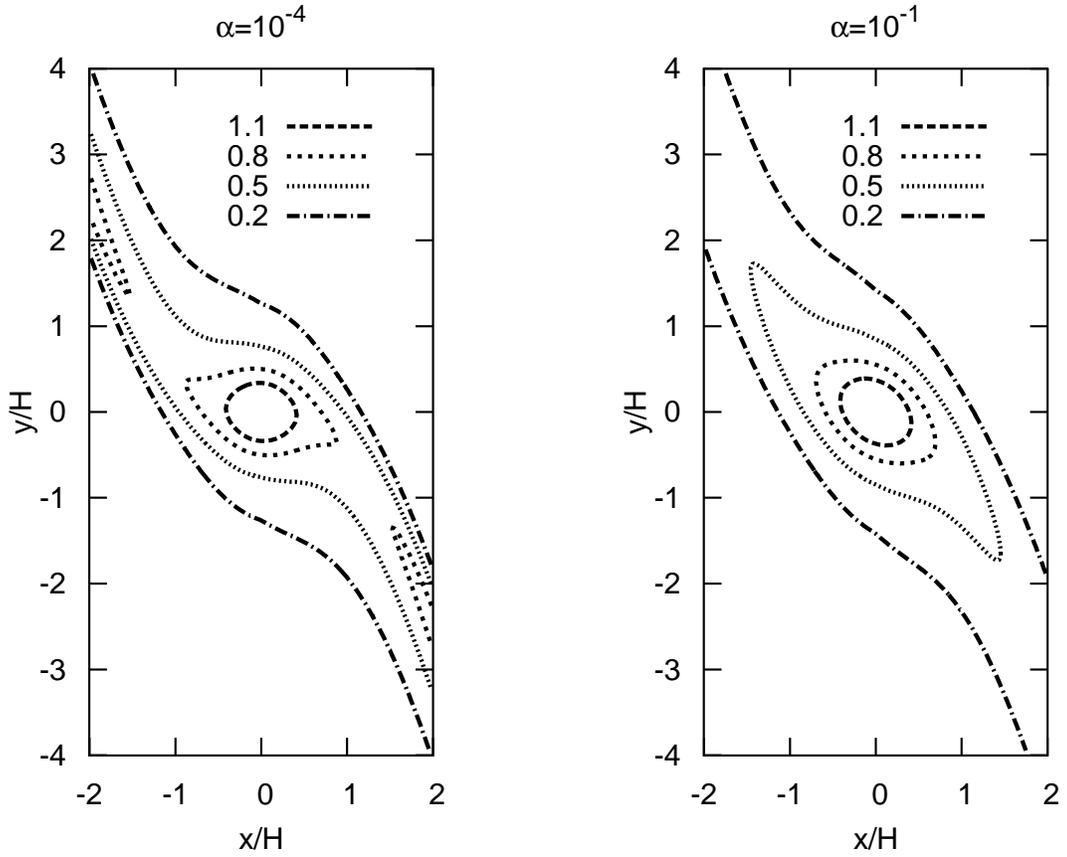}
 \caption{Contour plot of the density fluctuation 
 $\delta \rho/\rho_0$ around the planet
 (located at the origin) for 2D mode of 3D calculation with $L_z=2.5H$
 with $\alpha=10^{-4}$ (left) and $\alpha=10^{-1}$ (right).  
 Axisymmetric mode is not included.} 
 \label{fig:2D_dens}
\end{figure}

Since viscosity, or any form of dissipation, damps density contrast in
general, one may think that Lindblad torque is a decreasing function of
viscosity.
\footnote{  
The effect of viscosity on the fluid elements trapped in the
horseshoe regions enhances the resulting torque since viscosity keeps
the asymmetry of the potential vorticity.  This applies corotation
torque, see Masset (2001) for detail.
In shearing-sheet calculations presented here, however, there is no
corotation torque since background values are assumed to be constant.  
}
However, our result shows that one-sided torque peaks at 
$\alpha \sim 1$. 
This result
originates from two different effects of viscosity.  First, viscosity
damps spiral density wave, as discussed by Papaloizou and Lin (1984) or
Takeuchi et al. (1996). 
The second effect of viscosity may be observed by looking at the density
structure in the vicinity of the planet.  Figure
\ref{fig:2D_dens} shows the contour of density profile of the $xy$-plane
 for $\alpha=10^{-4}$ and $\alpha=10^{-1}$.  It is
clear that the oval density profile in the vicinity of the planet is
slightly tilted for large 
viscous coefficient, while the spiral density wave is damped.  This
tilt produces asymmetry of density structure in the $y$-direction,
thereby exerting one-sided torque on the planet.

\begin{figure}
 \plotone{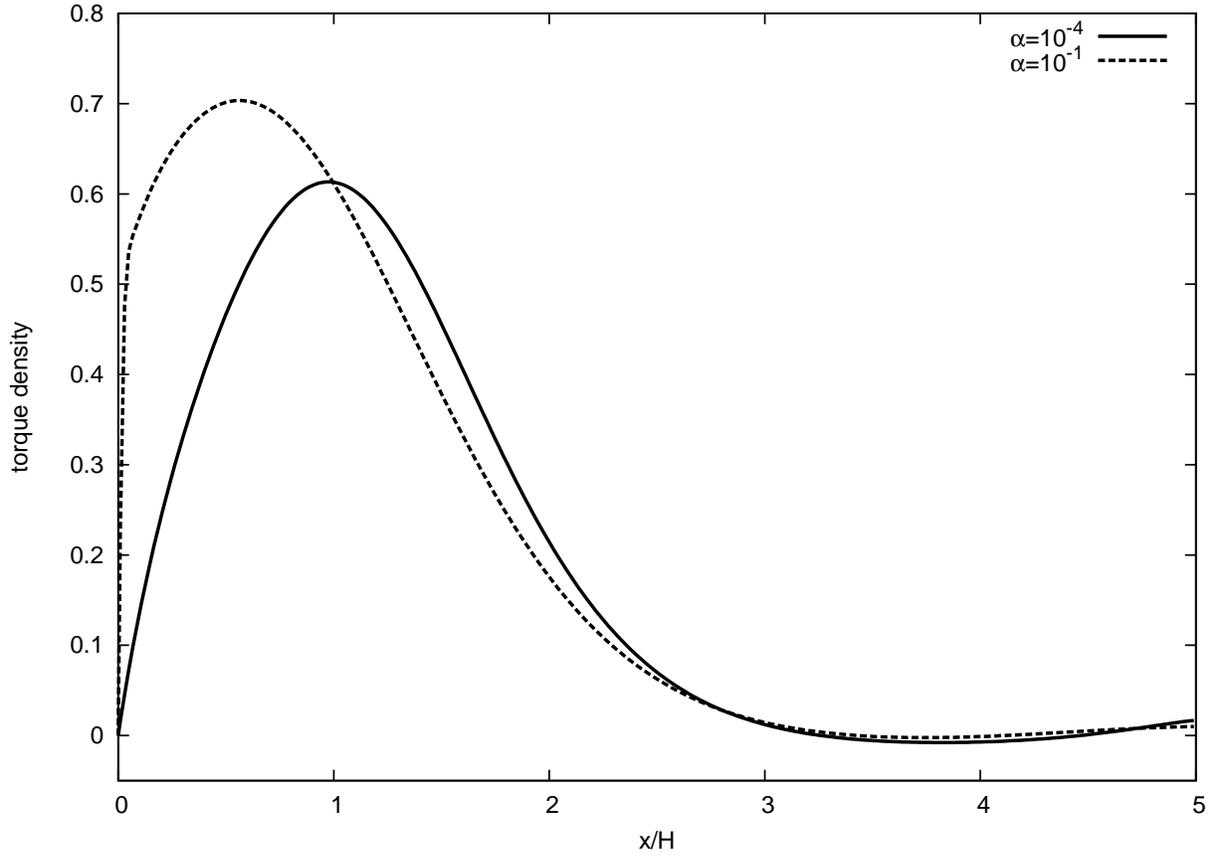}
 \caption{Torque density given by equation \eqref{def_tqdens} for
 $\alpha=10^{-4}$ (solid line) and $\alpha=10^{-1}$ (dashed line).
 Values are normalized according to equation 
 \eqref{norm_tqdens}.} 
 \label{fig:torque2D_compare}
\end{figure}

\begin{figure}
 \plotone{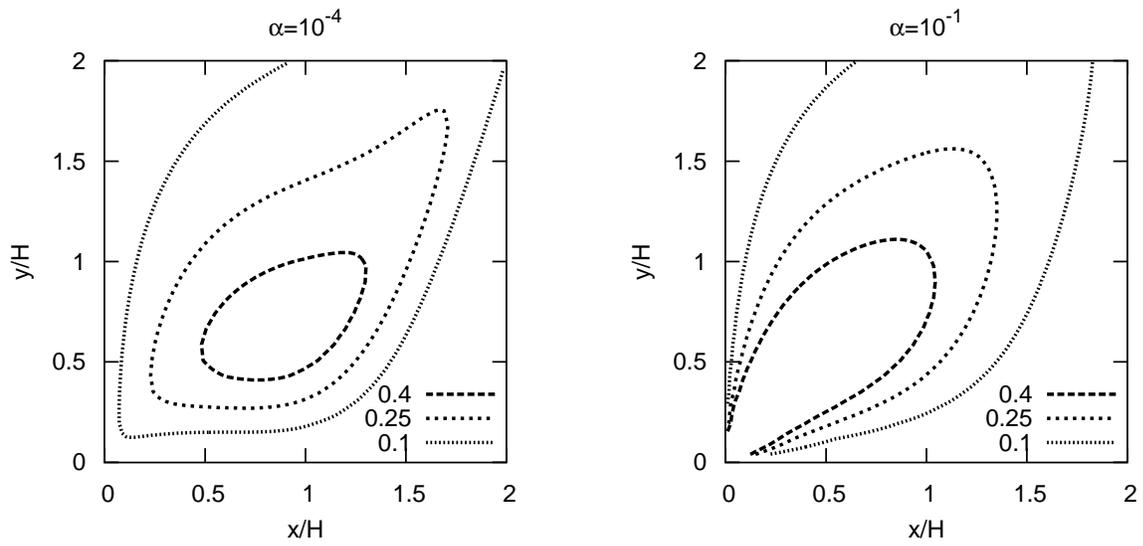}
 \caption{Forward-back asymmetry of the torque distribution given by
 equation \eqref{def_forwardback_torque} for $\alpha=10^{-4}$ and
 $\alpha=10^{-1}$.}  
 \label{fig:forwardback_2D}
\end{figure}

Figure \ref{fig:torque2D_compare} shows the
torque density for $\alpha=10^{-1}$ and
$\alpha=10^{-4}$, respectively.  It is clear that the peak of torque
density locates slightly closer to the planet for $\alpha=10^{-1}$ than
 $\alpha=10^{-4}$. 
The main contribution to the torque comes from the flow
structure in the vicinity of the planet, which is evident when we
consider the asymmetry of the density perturbation between $y>0$ and
$y<0$ regions.  In Figure \ref{fig:forwardback_2D}, we plot the contour
for the sum of torque at $y>0$ and $y<0$ region,
\begin{equation}
 T(x,y) + T(x,-y),
  \label{def_forwardback_torque}
\end{equation}
as a function of $x$ and $y(>0)$.  This shows the asymmetry of torque
distribution in the $y$-directions.  In Figure \ref{fig:forwardback_2D}, 
we compare the results for two different values of viscosity,  
 $\alpha=10^{-4}$ and $\alpha=10^{-1}$.  In the case of
$\alpha=10^{-4}$, the forward-back difference of the torque in the
vicinity of the planet almost vanishes since the density perturbation
around the planet is symmetric in the $y$-direction.  In the case of
$\alpha=10^{-1}$, however, the most of the contribution to the torque
comes from the region very close to the planet since the oval density
perturbation structure is tilted in $y<0$ direction (see also Figure
\ref{fig:2D_dens}).

\subsection{3D Calculation and the Effects of Smoothing Length}
\label{results:3D}

\begin{figure}
 \plotone{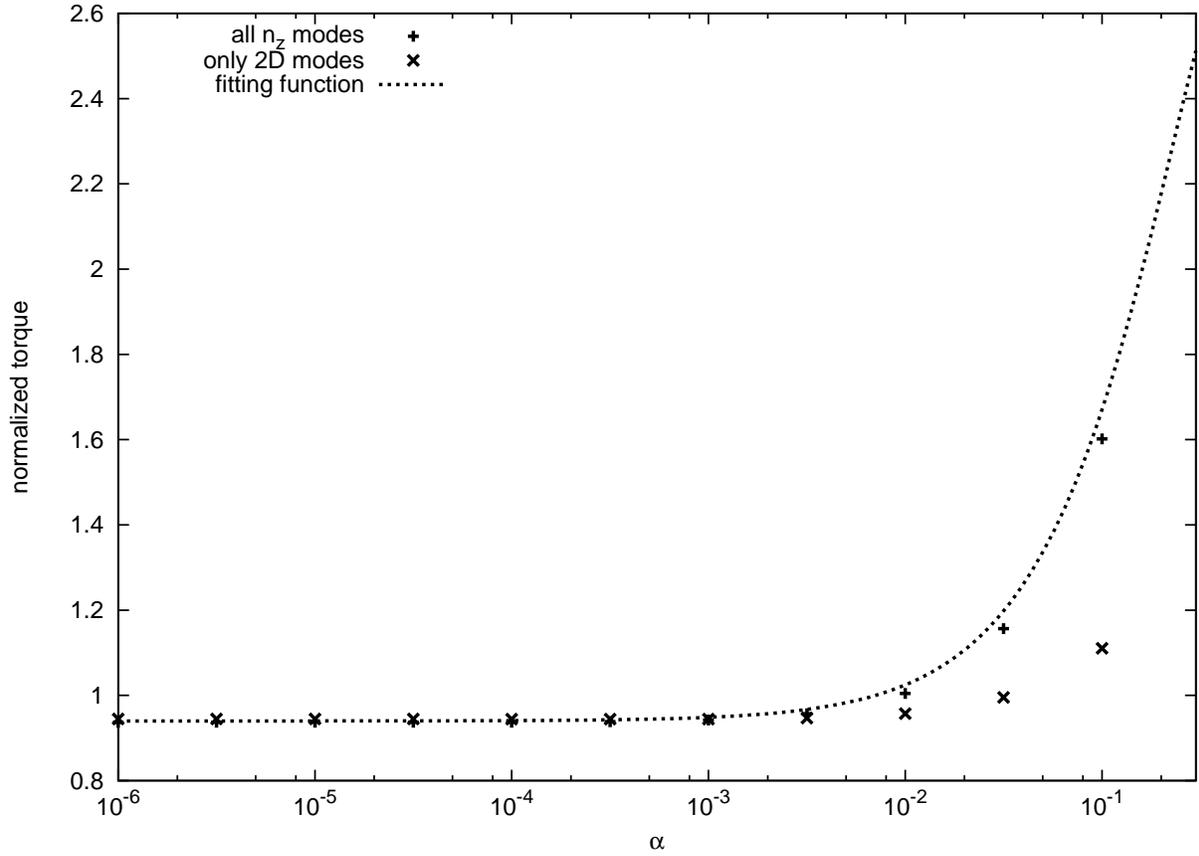}
 \caption{Variation of the torque as a function of viscosity parameter
 $\alpha$ obtained for 3D calculations (plus).  For comparison, we plot
 the results restricted to 2D modes by cross.  Dashed line shows the
 fitting function given by equation \eqref{torque_fit}.} 
 \label{fig:3D_torque}
\end{figure}

In this section, we present the results of 3D calculation.  Figure
\ref{fig:3D_torque} shows the one-sided torque as a function of
viscosity.  The qualitative behavior of torque increasing with viscosity
is unchanged, but the enhancement of torque becomes large compared to 2D
calculation.  We fit the data and obtain the following empirical relation
\begin{equation}
 \dfrac{T}{\rho_0 r_{\rm p} H^2 c^2} 
  = ( 0.94+10 \alpha ) e^{-1.5 \alpha} 
  \label{torque_fit}
\end{equation}
for $\alpha<0.3$ and $L_z=2.5 H$.  
We have chosen a form of fitting function in such a way that torque
converges to a non-zero value for $\alpha \ll 1$, peaks at 
$\alpha \sim 1$, and decreases to zero for $\alpha \gg 1$.  
This fitting formula is in reasonably
good agreement with our calculation in this range of viscosity
coefficient.

\begin{figure}
 \plotone{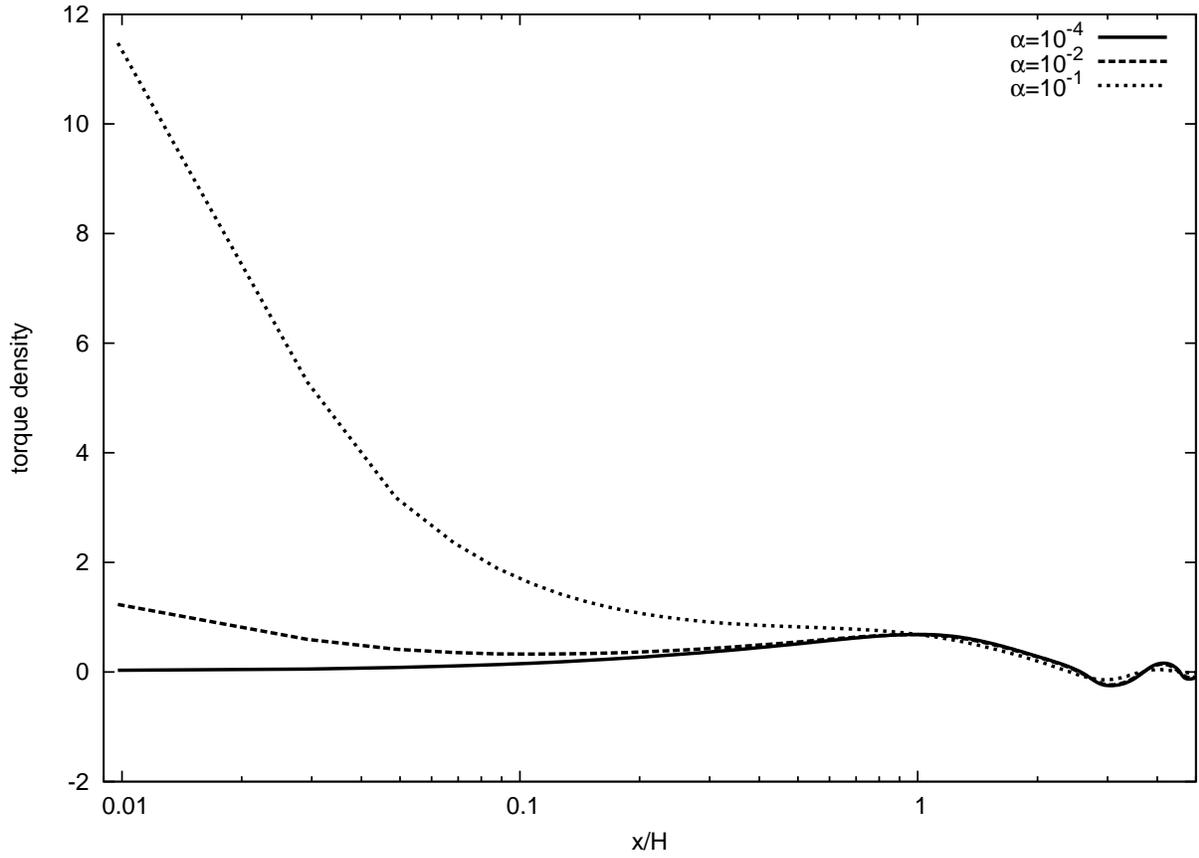}
 \caption{Torque density profile obtained by 3D calculation with various
 $\alpha$.  Solid line shows the results of $\alpha=10^{-4}$, thick
 dashed line $\alpha=10^{-2}$, and thick dotted line $\alpha=10^{-1}$.} 
 \label{fig:tqdens_3D}
\end{figure}

\begin{figure}
 \plotone{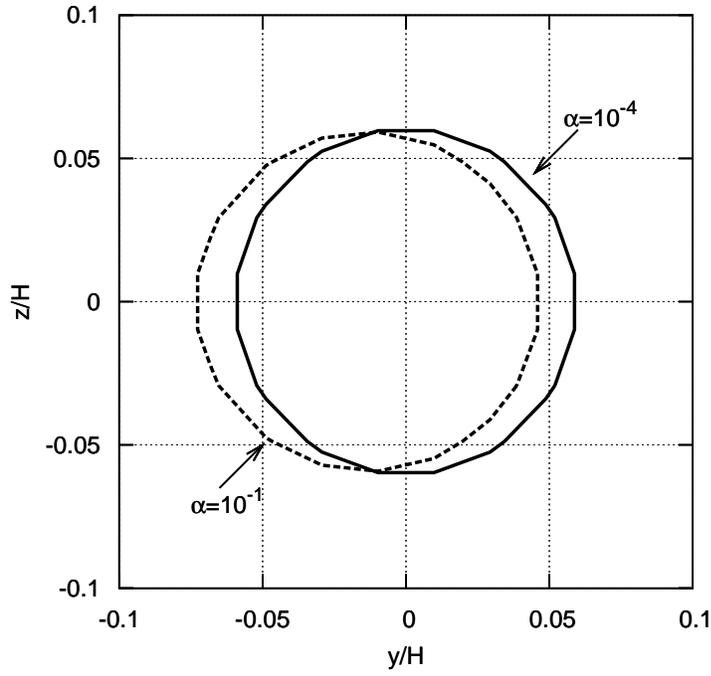}
 \caption{The contour plot of 3D mode density structure in $yz$-plane
 at $x=0.068H$ with
 $\alpha=10^{-4}$ (solid line) and $\alpha=10^{-1}$ (dashed line).  The
 lines show the contours of $\delta \rho/\rho_0=10$.} 
 \label{fig:3D_dens_planet}
\end{figure}

Just as in the calculations restricted to 2D modes, 
the density structure mainly 
in the vicinity of the planet contributes to the one-sided torque
if large values of viscosity is assumed.  
Figure
\ref{fig:tqdens_3D} shows the torque density profile obtained for
different viscosity coefficients.  The more viscous is the
disk, the closer to the planet is the location of the dominant
contribution to the torque.
Figure \ref{fig:3D_dens_planet} compares the density 
structure in the $yz$-plane 
at $x=0.068H$ for calculations with $\alpha=10^{-4}$ and
$\alpha=10^{-1}$.  
The asymmetry in the $y$-direction
in strongly perturbed region is present in calculations with large
viscosity.

In 3D calculations, gravitational potential
in the vicinity of the planet 
is not as strongly softened as in 2D modes.
In the vicinity of the planet, gas feels the gravitational potential
that decreases as $-1/r$, where $r$ is the distance from the planet,
in full 3D calculations.  However,  
the gravitational potential varies only logarithmically
when $r \lesssim H$ (see equation \eqref{psi_unstrat}) in calculations
restricted to 2D modes. 
 Since the density fluctuation around the planet is
given by $\delta \rho / \rho_0 \sim \psi_{\rm p}/c^2$, 
the deeper gravitational potential gives the higher value of the
torque exerted on the planet if there is a substantial asymmetry in the
density structure.

Calculations restricted to two-dimensional modes 
predict one-sided torque very well if viscosity parameter is small,
since regions close to the planet ($r \lesssim H$) 
is not very important.  
However, if large values of viscosity is used, it is necessary to
perform full three-dimensional calculations  
since the density structure close to the planet is
important.   

At the end of Section \ref{Lzsize}, we addressed that varying
smoothing length changes one-sided torque upto $30 \%$.  
Below, we discuss the effects of smoothing length in our
calculation and argue that the values of one-sided torque obtained for
large viscosity, $\alpha \gtrsim 0.01$, seems to be the lower limit of
the one-sided torque.

\begin{figure}
 \plotone{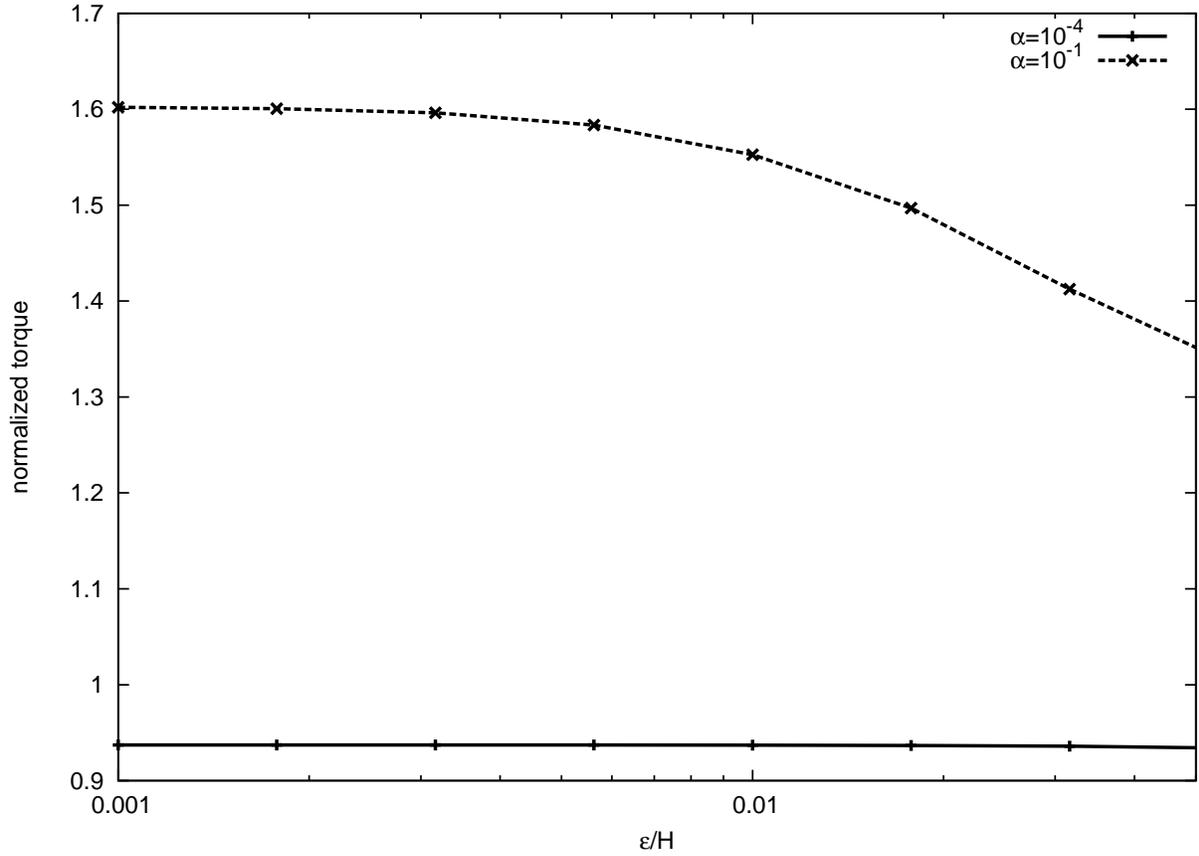}
 \caption{One-sided torque obtained by 3D calculation for different
 smoothing length.  Horizontal axis shows the values of smoothing length
 $\epsilon_{\rm 3D}$ and vertical axis shows the one-sided torque.
 Calculations with $\alpha=10^{-4}$ (solid line) and
 $10^{-1}$ (dashed line) are shown.}
 \label{fig:vareps}
\end{figure}

Figure \ref{fig:vareps} shows how one-sided torque of
three-dimensional calculation varies with smoothing length
$\epsilon_{\rm 3D}$.  Results with $\alpha=10^{-4}$ and
$\alpha=10^{-1}$ are shown.
In this calculation, we used the box size with
$(L_x,L_y,L_z)=(10H,10H,2.5H)$ and the grid number with
$(N_x,N_y,N_z)=(512,512,128)$.  Note that resolution in the
$y$-direction is better by a factor of four compared to the parameters
used in Figure \ref{fig:3D_torque}.  
Grid resolution is $\sim 0.02H$ in all directions in Figure
\ref{fig:vareps}.

It is shown that one-sided torque converges well
for small values of viscosity parameter.  If $\alpha$ is as large as
$0.1$, the calculation converges for sub-grid smoothing lengths, and 
results vary approximately $30 \%$ if smoothing length with twice the
grid scale ($\sim 0.04H$) is assumed. 
The qualitative behavior of one-sided torque can be understood if we
notice that smaller 
smoothing length gives a deeper potential in the vicinity of the
planet, and if large values of viscosity is assumed, density structure
in the vicinity of the planet is important for one-sided torque.

\begin{figure}
 \plotone{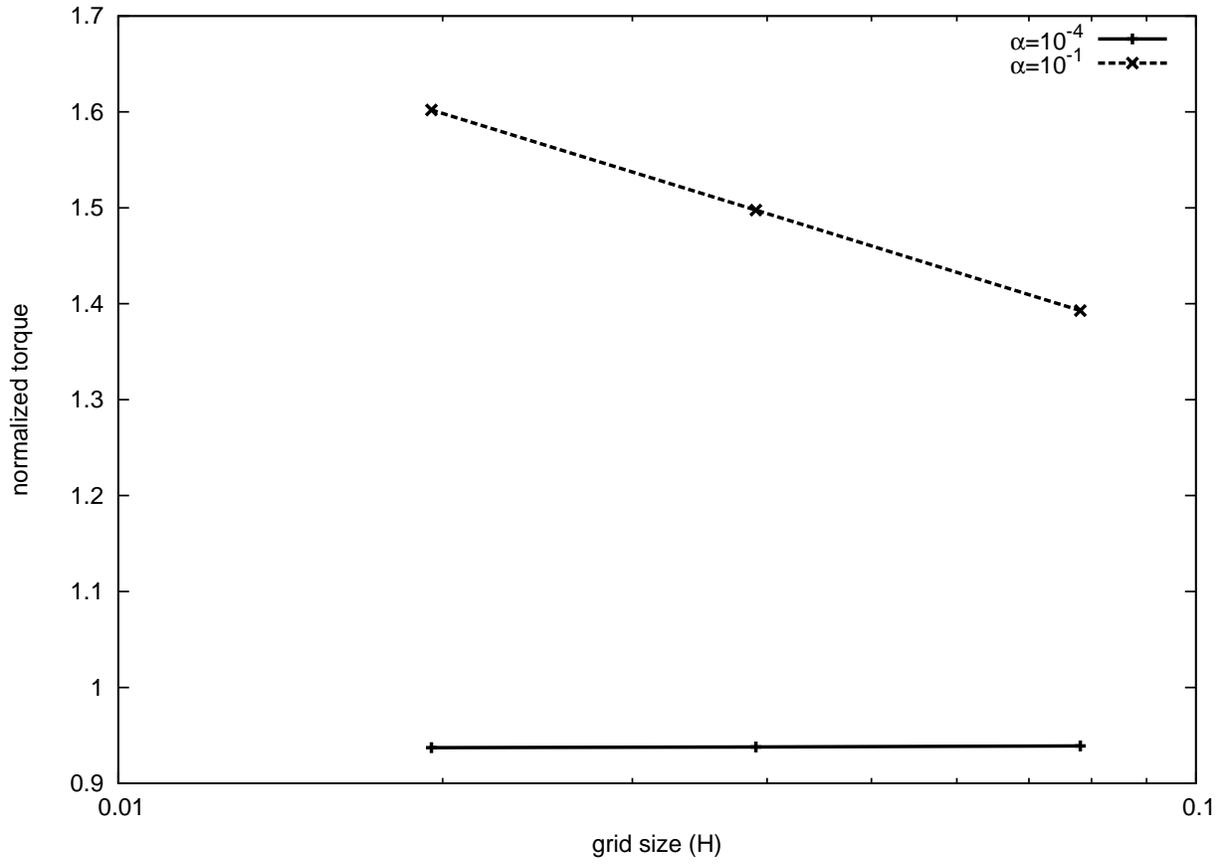}
 \caption{One-sided torque obtained by 3D calculation for different
 grid resolutions.  Horizontal axis shows the values of grid size (the
 same for all three dimensions) and vertical axis shows the one-sided
 torque.  Softening parameter $\epsilon_{\rm 3D}=10^{-3}H$ is used.  
 Calculations with $\alpha=10^{-4}$ (solid line) and
 $10^{-1}$ (dashed line) are shown.}
 \label{fig:varres}
\end{figure}

Since contribution to one-sided torque in case of large viscosity mainly
comes from the density structure close to the planet, it may depend on
the grid resolution used in the calculation.  Figure
\ref{fig:varres} shows the one-sided torque obtained for different
resolutions.  Calculations with $(N_x,N_y,N_z)=(512,512,128)$,
$(256,256,64)$, and $(128,128,32)$ are shown while keeping
$(L_x,L_y,L_z)=(10H,10H,2.5H)$.  These values corresponds to grid
resolutions with $\Delta x=\Delta y=\Delta z=0.02H$, $0.04H$, and
$0.08H$, respectively.  The value of softening parameter is kept
$\epsilon_{\rm 3D}=10^{-3}H$ for all the calculations.
It is shown that the value
of one-sided torque for $\alpha=10^{-4}$ 
is well converged while the
value of the torque for $\alpha=10^{-1}$ varies 
approximately $30 \%$ when grid resolution is
varied by a factor of four.  We note that in case of small viscosity, we 
have obtained well-converged values of one-sided torque since the
effective Lindblad resonances, which are located at
 $|x| \gtrsim (2/3)H$, are all well resolved. 

We note that the vertical averaging, large softening length, and
coarser grid all introduce more softened potential in the vicinity of
the planet and therefore one-sided torque becomes smaller in case of
large viscosity.  
Equations \eqref{psi_full3D} and \eqref{psi_unstrat} show
that the vertically averaged potential diverges at the location of the
planet more mildly.  Gravitational force becomes weaker in the vicinity
of the planet if we use larger values of softening length.  The
larger grid scale cuts off gravitational potential at larger distance.

From the behaviors when we vary the softening
parameter and the grid scale,  
we conclude that the values of the one-sided torque 
obtained in our calculation for 
large viscosity is the lower limit.  
We also argue 
that the numerical treatment in the vicinity of the planet may cause a
quantitative difference in the one-sided torque since it may change the 
density structure in the vicinity of the planet. 
If high values of viscosity is assumed, the density structure close to
the planet is important and therefore, three-dimensional
calculation with high resolution 
and small softening parameter is essential.  
The resolution and the softening parameter 
should be determined, in
principle, by considering the realistic size of the planet.

\section{Analytic Treatment of Density Structure around the Planet}
\label{analyticdiscussion}

We have seen that the viscosity exerted on the disk can change the
density structure in the vicinity of the planet and in a viscous disk,
the planet experiences one-sided torque from the gas well inside the
effective Lindblad resonance.  In this section, we show that a
dissipative force distorts the density structure close to the planet
by using a simple analytic model.

We consider a two-dimensional model with a friction force exerted only
in the $y$-direction.  We consider equations
\eqref{EoC_pert_four}-\eqref{EoMy_pert_four} with viscosity terms
replaced by friction.
  We assume
$k_x(t) \sim 0$ since significant excitation of perturbation occurs
when radial wavenumber becomes zero.  
We also set $k_z=0$ and consider 2D perturbation for simplicity.  
The set of equations we solve is 
\begin{equation}
 \dfrac{d \sigma}{dt} + ik_y \delta v_y = 0,
  \label{EoC_model}
\end{equation}
\begin{equation}
 \dfrac{d}{dt} \delta v_x - 2\Omega_{\rm p} \delta v_y = 0,
  \label{EoMx_model}
\end{equation}
and
\begin{equation}
 \dfrac{d}{dt} \delta v_y + \dfrac{1}{2} \Omega_{\rm p} \delta v_x 
  = - c^2 ik_y \left( \dfrac{\psi_{\rm p}}{c^2} + \sigma \right) -
  \gamma \delta v_y,
  \label{EoMy_model}
\end{equation}
where $\sigma=\delta \rho / \rho_0$ and $\gamma$ is a drag coefficient.

Using equations \eqref{EoC_model} and \eqref{EoMx_model} to eliminate
$\delta v_{y}$ and integrating once the resulting equation, we obtain
\begin{equation}
 \delta v_x = -\dfrac{2\Omega_{\rm p}}{i k_y} \sigma + V,
  \label{eq_oneint}
\end{equation}
where $V$ is a constant of integration. 
Physically, $V$ is the perturbation of vortensity, which is actually
conserved in this model since we neglect terms with $k_x(t)$ and we
assume friction force is exerted only in the $y$-direction.  Therefore,
$V$ is zero if we assume there is no vortensity perturbation initially.
Since we are interested in the perturbation that arises from the planet
potential, we assume $V=0$ hereafter.

Using equations \eqref{EoC_model} and \eqref{eq_oneint} to eliminate
$\delta v_x$ and $\delta v_y$ from equation \eqref{EoMy_model}, 
we obtain a single 
telegraph equation with a source terms,
\begin{equation}
 \dfrac{d^2 \sigma}{dt^2} + \gamma \dfrac{d\sigma}{dt} + s_0^2
  \sigma = -k_y^2 \psi_{\rm p},
  \label{telegraph}
\end{equation}
where
\begin{equation}
 s_0^2 = c^2 k_y^2 + \Omega_{\rm p}^2 .
  \label{properfreq}
\end{equation}
The right hand side of equation \eqref{telegraph} is the source of
perturbation that is induced by the planet's gravity.   
The source potential $\psi(k_x(t),k_y)$ depends on time through the
time-dependence of $k_x(t)$.

Let us now derive the steady state profile of density perturbation in
$(t,x,y)$-coordinate space.
In order to solve equation \eqref{telegraph}, we Fourier transform
perturbation in $t$-direction
\begin{equation}
 f(t,k_y) = \int_{-\infty}^{\infty} ds f(s,k_y) 
  e^{i s t}, 
  \label{FT_t}
\end{equation}
where $f$ denotes any of perturbation quantities.
We note that the Fourier transformation in $t$-direction is equivalent
to the inverse Fourier transformation of $k_x$ modes into $x$-coordinate
space.  The ``frequency'', $s$, in equation \eqref{FT_t} is the
frequency of perturbation experienced by a single mode specified by
$k_y$. 

The real space quantity $f(x,y)$ is obtained by
\begin{equation}
 f(x,y) = \int_{-\infty}^{\infty} dk_y \int_{-\infty}^{\infty} dk_x
  f(k_x,k_y) e^{i(k_x x + k_y y)}.
  \label{inverseFT_start}
\end{equation}
In a steady state, the value of $k_x^{\prime}$ can be taken arbitrary
[see discussion in Section \ref{linear_anal}].  Therefore, we can take
$k_x^{\prime}=0$ without loss of generality.  Time-dependence of $k_x$
is given by
\begin{equation}
 k_x(t) = \dfrac{3}{2} \Omega_{\rm p} k_y t,
\end{equation}
and therefore, we obtain
\begin{equation}
 d k_x = \dfrac{3}{2} \Omega_{\rm p} k_y dt.
\end{equation}
Changing the integration variable from $k_x$ to $t$ in equation
\eqref{inverseFT_start}, we obtain
\begin{eqnarray}
& f(x,y) =& \int_{0}^{\infty} dk_y \int_{-\infty}^{\infty} dt \dfrac{3}{2}
  \Omega_{\rm p} k_y f(t,k_y) \exp\left[ \dfrac{3}{2} k_y x \Omega_{\rm
				   p} t + k_y y \right]  \nonumber \\
&&  - \int_{-\infty}^{0} dk_y \int_{-\infty}^{\infty} dt \dfrac{3}{2}
  \Omega_{\rm p} k_y f(t,k_y) \exp\left[ \dfrac{3}{2} k_y x \Omega_{\rm
				   p} t + k_y y \right] .
  \label{FT_tky}
\end{eqnarray}
The value $f(t,k_y)$ appeared in this equation is Fourier transformed in
the $t$-direction according to equation \eqref{FT_t}.  Substituting
equation \eqref{FT_tky} into \eqref{FT_t}, performing integral in $t$
and $s$, we obtain the relationship between $f(s, k_y)$ and $f(x,y)$,
\begin{eqnarray}
& f(x,y) = &2\pi \int_{0}^{\infty} dk_y \dfrac{3}{2} \Omega_{\rm p} k_y 
  \nonumber \\
&&  \times \left[ e^{ik_y y} f(s=-(3/2)\Omega_{\rm p}k_y x, k_y)
   + e^{-ik_y y} f(s=(3/2)\Omega_{\rm p}k_y x, -k_y)
  \right] .
\label{inverseFT}
\end{eqnarray}

One-sided torque is calculated by equation \eqref{def_totaltorque}
(but there is no integral in the $z$-direction in two-dimensional
problem considered in this section).  
We choose the phase of the potential such that 
$\psi_{\rm p}(s, k_y)$ is real.  It is possible because the potential of
the planet is spherically symmetric and the planet is fixed at the
origin of the coordinate system.  Thus, we obtain
\begin{equation}
 T \propto \int_{0}^{\infty} dk_y k_y^2 \int_{0}^{\infty} ds \mathrm{Im}
  \left[ \sigma(s,-k_y) \right] \psi_{\rm p}(s,-k_y), 
  \label{torque_xky}
\end{equation}
where constant of proportionality does not depend on  $k_y$, $\gamma$, 
and $x$ and ``$\mathrm{Im}$'' denotes the imaginary part.
From this equation,  
torque is exerted on the planet only when $\sigma(s,k_y)$ and
$\psi_{\rm p}(s,k_y)$ are out of phase.  In other words, the amount
of torque can be estimated by looking at the imaginary part of
$\sigma(s,k_y)$, provided that the phase of Fourier transform is
taken in such a way that $\psi_{\rm p}(s,k_y)$ is real.  
If density perturbation and the
potential are in phase, the $y$-component of the force exerted on the
planet is symmetric in the $y$-direction and there is no net one-sided
torque when integrated over $y$.

Now we consider the solution of equation \eqref{telegraph}.  By Fourier
transform in $t$-direction, we obtain
\begin{equation}
 \sigma (s,k_y) 
  = - \dfrac{(s_0^2 - s^2) - i s \gamma}{(s_0^2 -
  s^2)^2 + s^2 \gamma^2} k_y^2 \psi_{\rm p}(s,k_y) , 
  \label{telegraph_sol}
\end{equation}
and as equation \eqref{inverseFT} indicates, it is sufficient to
consider $s=(3/2)\Omega_{\rm p} k_y x$ mode.

In the limit of small
friction, $\gamma \to 0$, $\sigma$ and $\psi_{\rm p}$ are out of phase
only at the resonance, which is located at
\begin{equation}
 s^2 = \dfrac{9}{4} \Omega_{\rm p}^2 k_y^2 x^2 = s_0^2.
  \label{resonance_loc}
\end{equation}
From equation \eqref{properfreq}, we see that this corresponds to
the location of effective Lindblad resonance (Artymowicz 1993).  There
is a phase shift of $\pi$ only in the vicinity of the resonance and
torque is localized at the resonance location.  This localization of the
torque comes from our assumption of zero radial pressure term in
equation \eqref{EoMx_model}.  We note that in this case, the resulting
torque is independent of the amount of friction if integrated over the
resonance width (Meyer-Vernet and Sicardy 1987).
  This can be seen by considering the imaginary part of
the density fluctuation $\sigma$ limits to, for sufficiently small
$\gamma$, 
\begin{equation}
 \lim_{\gamma \to 0} \mathrm{Im}[\sigma(s,k_y)] = 
  \lim_{\gamma \to 0} \dfrac{s \gamma}{(s_0^2 - s^2)^2 +
  s^2 \gamma^2} k_y^2 \psi_{\rm p} = \pi \delta_{D}(s_0^2 -
  s^2) k_y^2 \psi_{\rm p},
\end{equation}
where $\delta_D(x)$ is 
Dirac's delta function.  The last expression is independent of the
amount of friction, and the torque exerted by the mode $k_y$ is
determined by the strength of the source $\psi_{\rm p}$ at the
resonance.   
 Using equations \eqref{properfreq} and
\eqref{resonance_loc}, we see that there is no resonance close to the
planet, or region $|x|<x_c$, where
\begin{equation}
 x_c = \dfrac{2}{3}\dfrac{c}{\Omega_{\rm p}}.
\end{equation}
In this region, density perturbation and source term are in phase.  In
terms of real space coordinate $(x,y)$, this corresponds to 
$\psi_{\rm p}(x,y) \propto \sigma(x,y)$.

In the case of finite friction, the situation is different.  
From equation \eqref{telegraph_sol}, we see that 
resonance width is given by $|s - s_0| \sim \gamma$.  
Therefore, when $\gamma \sim \Omega_{\rm p}$, the region in which
density perturbation and source term are out of phase can overlap the
location of the planet.  Note that $s_0$ is a function of $k_y$ and
$x$ [see equation \eqref{resonance_loc}], so ``the width of the
resonance'' refers to the width in the $x$-direction, and this width
varies with the mode in the $y$-direction ($k_y$) we consider.   
The finite width of the resonance causes the asymmetry in density
perturbation in $y>0$ and $y<0$ regions even at the location of the
planet, $x \sim 0$.  Since gravitational force exerted by the planet on
the disk is large in the vicinity of the planet, the significant amount
of one-sided torque can be exerted on the planet.
 
The amplitude of perturbation is suppressed when significant
friction is exerted.  From equation \eqref{telegraph_sol}, we see that, 
in the vicinity of the planet
\begin{equation}
 |\sigma| \sim \dfrac{1}{\sqrt{s_0^4 + s^2 \gamma^2}} k_y^2
  \psi_{\rm p}(s,k_y).
\end{equation}
The suppression is significant only when $\gamma$ exceeds
$\Omega_{\rm p}$. 

One-sided torque is calculated by equation \eqref{def_totaltorque}
(but there is no integral in the $z$-direction in two-dimensional
problem considered in this section).  
Using equations \eqref{torque_xky} and \eqref{telegraph_sol}, it is
possible to obtain 
\begin{equation}
T \propto \int_{0}^{\infty} dk_y k_y^4 \int_{0}^{\infty} ds 
  \dfrac{s \gamma}{(s_0^2 - s^2)^2 + s^2 \gamma^2}
  \psi_{\rm p}^2(s, -k_y),
  \label{torque_anal_int_before}
\end{equation}
where constant of proportionality does not depend on $k_y$, $\gamma$,
and $x$.  The value of the torque is determined by the competition
between the suppression of the amount of the density perturbation and
the amplification of the amount of the torque as the resonance width
becomes wider.  Therefore, it is expected that the torque peaks at
$\gamma \sim \Omega_{\rm p}$.  This explains the peak of the torque
for $\alpha \sim 1$ in the calculation of viscous disk presented in the
previous section.

The qualitative behavior of one-sided torque can be captured more
clearly if we 
consider a further simplified case.  We consider the case when the
forcing potential $\psi_{\rm p}(s, k_y)$ does not depend on
$s$,
\begin{equation}
 \psi_{\rm p} (s, k_y) = \Psi .
\end{equation}
This corresponds to the case where forcing potential is constant in the
$x$-direction.  
  In this case, it is 
possible to perform $s$ integral analytically,
\begin{eqnarray}
 \label{torque_anal_int}
 T \propto \int_0^{\infty} dk_y k_y^4 \times 
  \left\{
  \begin{array}{cc}
   \dfrac{1}{2\sqrt{4s_0^2 - \gamma^2}} 
    \left( \pi + 2 \tan^{-1} \left[ \dfrac{2s_0^2 - \gamma^2}{\gamma
			      \sqrt{4s_0^2 - \gamma^2}} \right]
    \right) 
    & \gamma^2 < 4s_0^2 \\[15pt]
   \dfrac{1}{2\sqrt{\gamma^2 - 4s_0^2}} \log 
   \left[ \dfrac{\gamma^2 - 2s_0^2 + \gamma \sqrt{\gamma^2 -
    4s_0^2}}{\gamma^2 - 2s_0^2 - \gamma \sqrt{\gamma^2 -
    4s_0^2}} 
   \right] 
   & \gamma^2 > 4s_0^2
  \end{array}
  \right.
\end{eqnarray}
If there is only one $k_y$ mode in the potential, this gives the exact
amount of torque. 
Otherwise, this gives the amount of torque exerted by each mode of
$k_y$.  Limiting values of equation \eqref{torque_anal_int} are
\begin{eqnarray}
 \begin{array}{cc}
 T \to \dfrac{\pi k_y^4}{2s_0} & \gamma \to 0  \\[15pt]
 T \propto \dfrac{1}{|\gamma - 2s_0|} & \gamma \sim s_0
  \\[15pt]
 T \to 0 & \gamma \to \infty .
 \end{array}
\end{eqnarray} 
Therefore, the amount of torque is independent of the amount of
dissipation in the limit of $\gamma \to 0$, it increases as $\gamma$
increases and peaks at $\gamma \sim s_0$, and then it decreases to
zero as $\gamma$ becomes very large.  In case of small $\gamma$, the
integrand of \eqref{torque_anal_int_before} is localized at 
$s \sim s_0$.  When $\gamma$ is large, however, it is
necessary to consider contribution from all the region of $s$. 
In general, since the amplitude of
forcing term $\psi_{\rm p}(s, k_y)$ increases as $s \to 0$, 
it is expected that the value of torque peaks at 
$\gamma \sim s_0$.  If the amplitude of forcing term cuts off at
$\Omega_{\rm p}^2 \sim c^2 k_y^2$, the peak of the torque is expected at
$\gamma\sim \Omega_{\rm p}$.

\section{Summary and Discussion}
\label{discussionsummary}

\subsection{Summary of Our Results}

In this paper, we have performed linear analyses of density
 structure as 
 a result of gravitational interaction between a planet and a viscous
 gas disk using shearing-sheet approximation.  We have obtained the
 density structure in the vicinity of the planet and discussed its
 effect on type I migration by calculating one-sided torque.
We have used time-dependent methods to obtain the density structure and
 a wide range of viscous parameter $\alpha$ is investigated.   
We have calculated one-sided torque, which is the torque exerted on the
 planet by the gas on one side of the planet's orbit.  One-sided torque
 converges to a well-defined value if viscosity is small, but increases
 as we increase the value of viscosity for $\alpha\gtrsim 0.01$, and
 then decreases as $\alpha$ exceeds the order of unity.  The values of
 one-sided torque we have obtained for large viscosity may be the lower
 limit of the actual value.  
 We note that total torque is not actually calculated in this paper
 since it is not possible within the shearing-sheet formalism.  However,
 we also emphasize that the results of shearing-sheet calculations are
 useful in calculating differential torque (Tanaka et al. 2002). 
We have seen that the torque exerted on the planet in a
viscous disk is primarily determined by the tilted spherical (or
spheroidal) density structure in  
the vicinity of the planet, in contrast to the inviscid case, where
main contribution of the torque comes from effective Lindblad resonances
and the horseshoe region.  In an inviscid disk, although the value of
density perturbation is large in the vicinity of the resonance, it is
symmetric in $y$-direction and therefore, torque exerted from $y>0$ and
$y<0$ cancels each other.  In a viscous disk, there is an asymmetry in
$y$-direction, resulting in increasing torque with increasing
viscosity until $\alpha \sim 1$.
 By using a simplified analytic calculations in Section
 \ref{analyticdiscussion}, we have shown that  
  it is possible to consider these effects of viscosity
 as the increase of the width of Lindblad resonances.   
 When the viscous coefficient is larger than 
$\alpha \gtrsim 1$, however, the torque becomes small with increasing
 viscous 
coefficient since the density perturbation is damped due to large
viscosity.  The torque can be a factor of two larger than inviscid
 case.
  The impact of detailed structure around the planet on type I migration
  has not yet been fully analyzed in previous numerical simulations
  probably because  
 relatively large values of viscosity ($\alpha \gtrsim 10^{-2}-10^{-1}$)
  are required to see this effect.  
Below, we show some discussions and future prospects.

\subsection{Prospects of Modified Local Analysis and Global Calculation}

\subsubsection{Validity of Linear Analysis on One-sided Torque}

In this subsection, we discuss whether or not linear
approximation well describes the real density structure around the
planet.
In the linear approximation,
the magnitude of perturbed quantities such as $\delta \rho / \rho_0$ or 
$\delta \vect{v}/c$ must be smaller than the order of unity.  In our
calculation, all the non-dimensional perturbed quantity is proportional
to the value of $GM_{\rm p}/Hc^2$, which is about $10^{-1}-10^{-2}$ for
typical values of protoplanetary nebula.  Therefore, 
when $\delta \rho / \rho$ becomes the
order of ten in the normalized calculations given in this paper, linear
approximation becomes less accurate.  Since the  
contribution of the torque in a viscous disk mainly comes from such
strongly perturbed regions, it may be necessary to perform high
resolution numerical calculation in order to fully obtain the
structure of the density fluctuation.  
  This is one of the future
prospects of this study.  However, we expect that the basic physics and
the order of magnitude of one-sided torque is similar to the linear
calculation.

\subsubsection{Differential Lindblad Torque}

The total torque that is exerted on the planet is the differential
torque, which is the difference of the torque exerted by the inner disk
($x<0$) and outer disk ($x>0$).  In the shearing-sheet calculations, the
inner and outer disks are symmetric and therefore torque exerted on the
two regions cancel each other.
We have seen that the one-sided torque is mainly exerted by the tilted
spheroidal density profile around the planet if strong viscosity is
exerted on the disk.  Therefore, we expect that
the asymmetry between the inner and outer disk comes from the asymmetry
of the density structure around the planet, probably within, or around,
the distance of the 
order of Bondi radius, $r_{\rm B} = GM_{\rm p}/c^2$.
Assuming that the differential torque $T_{\rm diff}$
mainly comes from the effects of curvature, the order of the magnitude
is 
\begin{equation}
 T_{\rm diff} \sim T_{\rm one-side} \times \dfrac{\delta x}{r_{\rm p}},
\end{equation}
where $T_{\rm one-side}$ is one-sided torque obtained in this paper, and
$\delta x$ is the length scale where most contribution to the torque
comes from.  In the inviscid case, $\delta x \sim H$ and therefore
\begin{equation}
 T_{\rm diff, inviscid} \sim T_{\rm one-side, inviscid} \times
  \dfrac{H}{r_{\rm p}} .
\end{equation}
In the viscous
case presented in this paper, $\delta x$ may be as small as Bondi
radius $r_{\rm B}$.  Therefore, the lower limit of the differential
torque in viscous disk may be estimated as
\begin{equation}
 T_{\rm diff, viscous} \sim T_{\rm one-side, viscous} \times
  \dfrac{r_{\rm B}}{r_{\rm p}},
\end{equation}
and the ratio of differential torque in viscous disk and that in
inviscid disk may be given
by 
\begin{equation}
 \dfrac{T_{\rm diff, viscous}}{T_{\rm diff, inviscid}}
  \sim \dfrac{T_{\rm one-side, viscous}}{T_{\rm one-side, inviscid}}
  \times \dfrac{r_{\rm B}}{H}.
\end{equation}
Since
\begin{equation}
 \dfrac{r_{\rm B}}{H} \sim \dfrac{M_{\rm p}}{M_{\rm c}}
  \left( \dfrac{r_{\rm p}}{H} \right)^3 
  = 8 \times 10^{-3} \left( \dfrac{10^{-6}}{M_{\rm p}/M_{\rm c}} \right)
  \left( \dfrac{0.05}{H/r_{\rm p}} \right)^{-3},
\end{equation}
and $T_{\rm one-side, viscous}/T_{\rm one-side, inviscid} \sim 2$ from
our calculation, differential torque in a viscous disk can be much
smaller than that in an inviscid disk, although one-sided torque becomes
larger when the value of viscosity is $\alpha \sim 1$.

  In order to include the effect of
asymmetry, which comes from curvature effect, it is necessary to proceed
to modified local approximation (Tanaka et al. 2002) or to perform a
global calculation.  In order to find the precise value of the torque
exerted on the disk and find how various physical processes change the
nature of type I migration, the time-dependent formulation of modified
local approximation is necessary.

\subsubsection{Corotation Torque}

In the shearing-sheet formalism, we cannot calculate 
corotation torque.  
For two-dimensional modes, this is because  all the background
quantities (and therefore vortensity) are assumed to be constant. 
For three-dimensional modes,  we note that, in Tanaka et al. (2002),
there is a singularity at corotation  
modes (see equation (A5) of their paper).  However, this singularity
does not account for the torque in itself.  Results of shearing-sheet
calculations must be combined with the higher order solutions of
modified local approximation in order to
find the torque exerted at corotation resonance, see equation (57) of
Tanaka et al. (2002).

If we assume that time derivatives of equations \eqref{EoC_pert} and
\eqref{EoM_pert} are zero in order to obtain a stationary solution, we
also find a similar 
singularity at $x=0$ for three-dimensional waves ($k_z \neq 0$), even we
have assumed the constant background density in the $z$-direction.
However, within the shearing-sheet formalism, it does not give a torque
at the corotation.  Also, this singularity does not seem to play a major
role in our time-dependent calculations, since there is no singularity
in our formulation.

If we proceed to the modified local approximation, we expect that it is
possible to calculate the corotation torque in the linear regime.  
  Recently, Paardekooper and
Papaloizou (2009a) has shown that large viscosity will push the
corotation torque into a linear regime.  Therefore, not only from the
point of view of differential torque but also from the view of the
corotation torque, extension of our time-dependent methods to modified
local approximation is an interesting future work.   

In summary, in order to predict the precise value of the torque, it is
necessary to use a modified local approximation or perform global
calculations, which can take into account 
the curvature effect.  Non-linear analysis may be important since the
main contribution comes from the region in the vicinity of the planet
where density is strongly perturbed.  However, we believe that the basic
physics that exerts torque onto the planet can be captured by this
linear analysis and therefore, high-resolution study is necessary.  
 In global calculations, we also note that it is also
possible to investigate the effects of gas accretion onto the central
star on type I migration, which is always present in a viscous disk but
is not captured in our local calculations.

\subsection{Flow Structure in the Vicinity of the Planet} 
\label{discuss:streamline}

Recently, Paardekooper and Papaloizou (2009b) has shown an interesting
results regarding the flow structure close to the planet.  
In this section, we observe streamline of the flow in the vicinity of
the planet and compare our results with those previously studied.
We note that since we have not calculated axisymmetric ($k_y=0$) modes,
the flow structure we have obtained is incomplete.  However, we can 
still find some qualitative difference between inviscid and viscous
calculations.

\begin{figure}
 \plotone{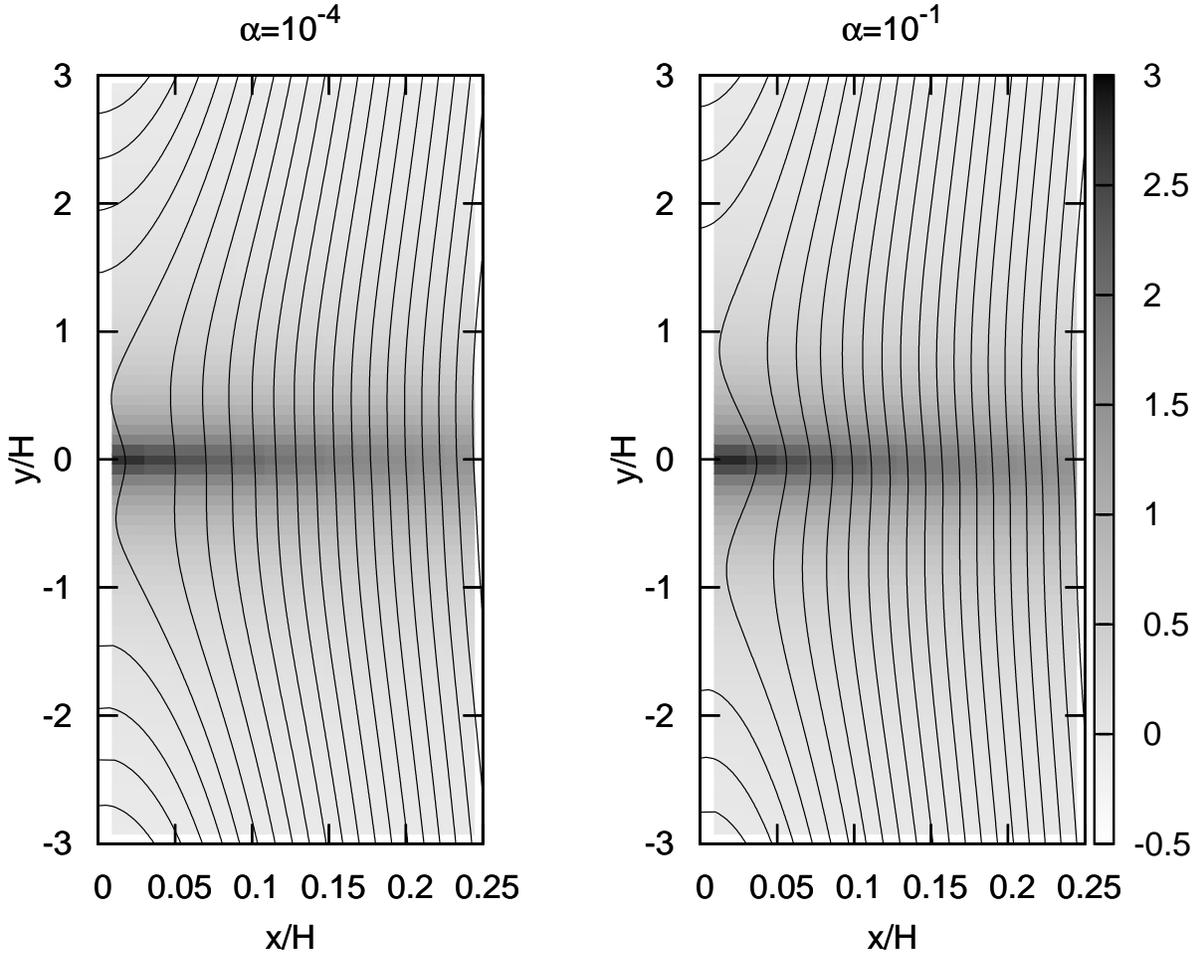}
 \caption{Density and streamline for calculations restricted to 2D mode
 in 3D calculation with $\alpha=10^{-4}$ (left) and $\alpha=10^{-1}$
 (right).  This corresponds to the
 calculations presented in Figure \ref{fig:2D_dens}.  Mass ratio between
 the central star $q=M_{\rm p}/M_{\rm c}$ 
 and disk aspect ratio $h=H/r_{p}$ are assumed in
 such a way that $q/h^3=0.0252$, which is the same as Paardekooper and
 Papaloizou (2009b).  Gray scale shows the density perturbation 
 $\delta \rho / \rho_0$ divided by $q/h^3$ and solid lines are
 streamlines. }
 \label{fig:dens_stream}
\end{figure}

Figure \ref{fig:dens_stream} shows the
streamlines plotted over the density fluctuations for calculations
corresponding to Figure \ref{fig:2D_dens}.  
Results with $\alpha=10^{-4}$ and $\alpha=10^{-1}$ are shown. 
Calculations are restricted to 2D modes of 3D calculation.
Mass ratio between the planet
and the central star $q=M_{\rm p}/M_{\rm c}$ and disk aspect ratio
$h=H/r_{\rm p}$ are assumed in such a way that $q/h^3=0.0252$, just as
Paardekooper and Papaloizou (2009b).

We find that the width of horseshoe orbit is $\sim 0.1H$ for
$\alpha=10^{-4}$, while the width becomes slightly narrower in
calculations with $\alpha=10^{-1}$.  We have obtained about a factor of
two smaller horseshoe width compared to Paardekooper and Papaloizou
(2009b) in calculations with small viscosity (see Figure 3 of their
paper.)  We think that this is probably because we have weaker
potential in the vicinity of the planet because of the effective
softening introduced by vertical averaging, see equation
\eqref{psi_unstrat}.  
In Paardekooper and Papaloizou (2009b), they also
obtained narrower horseshoe width in calculations with larger softening
parameters.  However, since we have neglected axisymmetric modes and
non-linear effects, which may be potentially important in determining
the streamline, this issue needs further investigation.

\subsection{Axisymmetric Modes}

Although we have neglected the axisymmetric modes in our calculations,
a viscous overstability in axisymmetric modes has been discussed,
especially in the context of Saturn's rings.  In this subsection, we
discuss whether this overstability affects our results.

Local linear analysis of viscous overstability using a simple
hydrodynamic model is performed by Schmidt et al. (2001).  Stability
criterion for isothermal case depends on the derivative of viscous
coefficient $\nu$ with respect to surface density perturbation.  We note
that the system of equations we have solved, equations
\eqref{EoC_pert_four}-\eqref{EoMz_pert_four}, are actually stable
against viscous overstability, since our assumption of viscosity
corresponds to $\beta=-1$ of Schmidt et al. (2001).
\footnote{
  Compare our
equations \eqref{EoC_pert} and \eqref{EoM_pert} and their equation
(8).  Note there is no self-gravity and bulk viscosity in our
calculations and we assumed isothermal equation of state.
}
  In this case, we
do not expect viscous overstability to occur and therefore,
we can safely neglect the axisymmetric modes.

However, disk may be prone to viscous overstability if different
prescription of viscosity is taken into account.  
If the non-linear consequence of
viscous overstability produces non-axisymmetric structure in the
vicinity of the planet, this may exert additional torque onto the
planet.  However, observation of the particle simulation performed by
Schmidt et al. (2001) (their Figure 1 for example) indicates that
non-axisymmetric modes are not strongly driven by the viscous
overstability and therefore this may not play an important role in
planetary migration.  Nonetheless, the effects of viscous overstability
seems to be an interesting extension of the analysis presented
here.

\subsection{Turbulent Disk}

In this subsection, we briefly discuss that our analysis may indicate
that the 
stochastic torque is important in a turbulent disk dominated by
magneto-rotational instability (see e.g., Nelson and Papaloizou 2004).
The effective turbulent viscosity originated in MRI 
may become as much as $\alpha \sim 0.1$ (e.g., Sano et 
al. 2004). 
Therefore, our calculation indicates that the density structure around
the planet is more important than the tidal wave launched at the
effective 
Lindblad resonances.  Since the density structure around the planet is
turbulent in the vicinity of the planet, stochastic force may be exerted
on the planet embedded in such a turbulent disk.

In order for the $\alpha$-viscosity prescription [see equation
\eqref{alpha_def}]
 to be a good
approximation of the turbulent flow, the length scale in consideration
must be larger than the eddy size.  In the present problem, the scale in 
consideration is smaller than disk thickness.  
In a turbulent flow driven by MRI, eddies of sizes on the order of
several tenths of the disk scale lengths exist.
This can be understood if we notice that the wavelength of the most
unstable mode is much smaller than the scale height of the disk with 
weak magnetic field.  If weak poloidal magnetic field,
$B_{z0}$, is exerted on the unperturbed disk, there always exists a net 
magnetic flux, 
\begin{equation}
 \langle B_z \rangle = B_{z0},
\end{equation}
where $\langle B_z \rangle$ is the horizontally averaged $z$-component
of magnetic 
field.  Therefore, there always exists eddies with scales of the
order of the most unstable mode of magneto-rotational instability with
net flux $B_{z0}$,
\begin{equation}
 \lambda \sim \dfrac{v_{A,0}}{\Omega},
\end{equation} 
where $\lambda$ is the eddy scale, 
 $v_{A,0} = B_{z0}/\sqrt{4\pi\rho}$
 is \alfven\  velocity constructed from the average poloidal field, and 
 $\Omega$ is Keplerian angular frequency.  In most cases, \alfven\
 velocity is smaller than sound speed by more than an order of a
 magnitude,
\begin{equation}
 v_{A,0} \lesssim 10^{-1} c.
\end{equation}
This indicates
\begin{equation}
 \lambda \lesssim 10^{-1} H.
\end{equation}
Thus, we can possibly use the effective
turbulent viscosity in order to study the qualitative effects of the
interaction between a turbulent disk and a planet.

However, the effective values of $\alpha$ may be obtained by
averaging turbulent stress over several scale heights, since it includes
all sizes of 
eddy that is important in turbulent flow driven by MRI.  Thus, the
use of $\alpha$ prescription may not be valid in discussing very small
scale structure, and full MHD calculation is necessary.  
  More detailed,
quantitative analyses require high-resolution numerical study of the
magnetized turbulent disk.

We have found that the effects of viscosity becomes apparent if $\alpha$
exceeds the value of $\sim 0.01-0.1$.   
Actual values of $\alpha$ in the disk is largely uncertain, but 
estimated values are around this value.
Numerical simulations by Sano et
al. (2004) indicates that the values of $\alpha$ varies from
$10^{-4}-10^{-1}$ depending on the setup.  
From the observational constraints of dwarf
novae, which seem to be better studied than protoplanetary disks,  it is
indicated that the values of $\alpha$ can vary from $0.01$ 
to $0.1$ (e.g., Cannizzo et al. 1988).  The values of $\alpha$ seems to
be an open issue for both theoretically and observationally.

\acknowledgments
Authors thank Taku Takeuchi and Hidekazu Tanaka for useful discussions.
We also thank the anonymous referee for a number of suggestions that
improved the paper.
This work was supported by the Grant-in-Aid for the Global COE Program
``The Next Generation of Physics, Spun from Universality and Emergence''
from the Ministry of Education, Culture, Sports, Science and Technology
(MEXT) of Japan.
The numerical calculations were carried out on Altix3700 BX2 at YITP in
Kyoto University. 
T. M. is supported by Grants-in-Aid for JSPS Fellows (19$\cdot$2409)
from MEXT of Japan.
S. I. is
supported by Grants-in-Aid (15740118, 16077202, and 18540238) from MEXT
of Japan.

\clearpage

\end{document}